\documentclass[12pt]{article}
\usepackage{amssymb,amsmath,amsfonts,amsbsy}
\usepackage{graphicx,verbatim}

\usepackage{epsf,wrapfig,epsfig}

\begin{document}

\title{Particle segregation by Stokes number for small neutrally buoyant spheres in a fluid}
\author{Phanindra Tallapragada and Shane D. Ross}

\maketitle

\begin{abstract}
It is a commonly observed phenomenon that spherical particles with inertia in an incompressible fluid do not behave as ideal tracers. Due to the inertia of the particle, the dynamics are described in a four dimensional phase space and thus can differ considerably from the ideal tracer dynamics. Using finite time Lyapunov exponents we compute the sensitivity of the final position of a particle with respect to its initial velocity, relative to the fluid and thus partition the relative velocity subspace at each point in configuration space. The computations are done at every point in the relative velocity subspace, thus giving a sensitivity field. The Stokes number being a measure of the independence of the particle from the underlying fluid flow, acts as a parameter in determining the variation in these partitions. We demonstrate how this partition framework can be used to segregate particles by Stokes number in a fluid. The fluid model used for demonstration is a two dimensional cellular flow. 
\end{abstract}

\section{Introduction}
It has long been observed that particles with a finite size and mass have different dynamics from the ambient fluid. Because of their inertia the particles do not evolve as point-like tracers in a fluid. This leads to preferential concentration, clustering and separation of particles as observed in numerous studies \cite{Segre, Tirumkudulu, Shinbrot}. The inertial dynamics of solid particles can have important implications in natural phenomena, e.g., the transport of pollutants and pathogenic spores in the atmosphere \cite{Isard, Schmale}, formation of rain clouds \cite{Facchini} by coalescence around dust particles and formation of plankton colonies in oceans \cite{Abraham}. Similarly, the inertial dynamics of reactant particles is important in the reaction kinetics and distribution of reactants in solution for coalescence type reactions \cite{Nishikawa}. Mixing sensitive reactions in the wake of bubbles has been shown to be driven by buoyancy effects of reactants \cite{Athanas}. Recently, a principle of asymmetric bifurcation of laminar flows was applied to the separation of particles by size and demonstrated the separation of flexible biological particles and the fractional distillation of blood \cite{Huang1, Huang2}. Innovative channel geometries have been empirically designed to focus randomly ordered inertial particles in microchannels \cite{DiCarlo}. These phenomena and related applications rely on the non-trivial dynamics of inertial particles in a fluid. In this paper, we demonstrate a theoretical tool to achieve particle segregation, by studying the sensitivity of the dynamics of inertial particles in a fluid.

We employ a simplified form of the Maxey-Riley equation \cite{Maxey} as the governing equation for the motion of inertial particles in a fluid. The dynamics of a single particle occur in a four dimensional phase space. The sensitive dependence of the particle motion on initial conditions is quantified using the finite time Lyapunov exponents (FTLE). It has been shown previously \cite{Shadden1, Shadden2}, that the ridges in the FTLE field act as separatrices. These are in general time dependent and go by the name of Lagrangian coherent structures (LCS).  We chose to do a simplified sensitivity analysis by perturbing the initial conditions in only two dimensions, in the initial relative velocity subspace.  We obtain a sensitivity field akin to a FTLE field but restricted to the relative velocity subspace and demonstrate numerically that the ridges in this field act as separatrices. The partitions in the relative velocity subspace created by these separatrices determine the eventual spatial distribution of particles in the fluid. Using this partitioning scheme we show how the Stokes number acts as a parameter in the separation of particles of different inertia or size. 

The paper is organized as follows. In \S 2 we review the the equation governing the inertial particle dynamics in a fluid and it's simplified form. In \S 3 we briefly review the background theory of phase space distributions of finite time Lyapunov exponents, which we use to quantify the sensitivity of the physical location of inertial particles with respect to perturbations in the initial relative velocity. 
We also describe our computational scheme to obtain the sensitivity field in the relative velocity subspace. 
In \S 4 we present results for the sensitivity field of the inertial particles in a cellular flow. 
In \S 5 we demonstrate our procedure for the segregation of particles by their Stokes number using the results from \S 4. 
In \S 6 we give numerical justification for the robustness of the sensitivity field to perturbations in the velocity field of the fluid. 
In \S 7 we discuss the results and give conclusions.

\section{Governing Equations}
Our starting point is Maxey and Riley's equation of motion of a rigid spherical particle in a fluid (\cite{Maxey}).
\begin{eqnarray}\label{eq:Maxey-Riley}
\rho_{p}\frac{\textit{d}\textbf{v}}{\textit{d}t} & = &\rho_{f}\frac{\textit{D}\textbf{u}}{\textit{D}t} +(\rho_{p}-\rho_{f})\textbf{g} 
- \frac{9\nu\rho_{f}}{2a^{2}} (\textbf{v} - \textbf{u} - \frac{a^{2}}{6}\nabla^{2}\textbf{u}) \\ \nonumber
& &-\rho_{f}(\frac{\textit{d}\textbf{v}}{\textit{d}t} - \frac{\textit{D}}{\textit{D}t}(\textbf{u} - \frac{a^{2}}{6}\nabla^{2}\textbf{u})) \\ \nonumber
& &-\frac{9\rho_{f}}{2a}\sqrt{\frac{\nu}{\pi}} \int_{0} ^{t}\frac{1}{\sqrt{t^{2}-\tau^{2}}}\frac{d}{d\tau}(\textbf{v} - \textbf{u} - \frac{a^{2}}{6}\nabla^{2}\textbf{u})d\tau
\end{eqnarray}
where \textbf{v} is the velocity of the solid spherical particle, \textbf{u} the velocity field of the fluid, $\rho_{p}$ the density of the particle, $\rho_{f}$, the density of the fluid, $\nu$ the kinematic of the viscosity of the fluid, $a$, the radius of the particle and \textbf{g} the acceleration due to gravity. The term on the right hand side are the force exerted by the undisturbed flow on the particle, the force of buoyancy, the Stokes drag, the added mass correction and the Basset-Boussinesq history force respectively. 
Eq \eqref{eq:Maxey-Riley} is valid under the following restrictions.
\begin{eqnarray}
a(\textbf{v} - \textbf{u} )/\textit{L} & << &  1 \\ \nonumber
a/L & << & 1 \\ \nonumber
(\frac{a^2}{\nu})(\frac{U}{L})& << &1
\end{eqnarray}\\
where $L$ and $U/L$ are the length scale and velocity gradient scale for the undisturbed fluid flow.
The derivative
\begin{equation}
\frac{D\textbf{u}}{Dt} = \frac{\partial\textbf{u}}{\partial t} + (\textbf{u}\cdot \nabla) \textbf{u}
\end{equation}
is the acceleration of a fluid particle along the fluid trajectory whereas the derivative
\begin{equation}
\frac{d\textbf{v}}{dt} = \frac{\partial\textbf{v}}{\partial t} + (\textbf{v}\cdot \nabla) \textbf{v}
\end{equation}
is the acceleration of a solid particle along the solid particle trajectory.

Eq.~\eqref{eq:Maxey-Riley} can be simplified by neglecting the Faxen correction and the Basset-Boussinesq terms \cite{Babiano}. We restrict our study to the case of neutrally buoyant particles, i.e $\rho_{p} = \rho_{f}$ . 
Writing \textbf{W} = (\textbf{v} - \textbf{u}), the relative velocity of the particle and the surrounding fluid, the evolution of \textbf{W} becomes
\begin{equation} \label{eq:dissipative1}
\frac{d\textbf{W}}{dt} = -(J+ \mu I) \cdot \textbf{W}
\end{equation}
and the change in the particle position is given by
\begin{equation}\label{eq:dissipative2}
\frac{d\textbf{x}}{dt} = \textbf{W} + \textbf{u}
\end{equation}
where $J$ is the gradient of the undisturbed velocity field of the fluid, $\textbf{u}$, and $\mu = \frac{2}{3}St^{-1}$ is a constant for a particle with a given Stokes number $St$.
Eqs \eqref{eq:dissipative1} and \eqref{eq:dissipative2} can be rewritten as the vector field
\begin{equation}\label{eq:dissipative}
\frac{d\textbf{x}}{dt} = \textbf{F}(\textbf{x})
\end{equation}
with $\textbf{x} = (\textbf{r},\textbf{W}) = (x,y,W_x,W_y) \in \mathbb{R}^4$. Eq \eqref{eq:dissipative} defines a dissipative system with constant divergence $- \frac{4}{3} \mu$. It has been shown by Haller \cite{Haller1} that an exponentially attracting slow manifold exists for general unsteady inertial particle motion as long as the particle Stokes number is small enough. For neutrally buoyant particles this attractor is $\textbf{W}=0$ (the \textit{xy} plane). Despite the global attractivity of the slow manifold, domains of instability exist in which particle trajectories diverge \cite{Babiano, Benczik, Haller2}.

\section{Sensitivity Analysis}

The Lyapunov characteristic exponent is widely used to quantify the sensitivity to initial conditions. A positive Lyapunov exponent is a good indicator of chaotic behavior. We have used the finite time version of the Lyapunov exponents, the FTLE, a measure of the the maximum stretching for a pair of phase points.

We review some important background regarding the FTLE below, following \cite{Shadden1, Shadden2, WigginsIde}.
The solution to eq \eqref{eq:dissipative} can be given by a flow map, $\phi_{t_{0}}^{t}$, which maps an initial point $\textbf{x}_{0}$ at time $t_{0}$ to $\textbf{x}_{t}$ at time $t$.
\begin{equation}
\textbf{x}_{t} = \boldsymbol{\phi} _{t_{0}}^{t}(\textbf{x}_{0}) 
\end{equation}
The evolution over a time $T$ of the displacement between two initially close phase points, $\textbf{x}(t_{0})$ and $\textbf{x}(t_{0})+ \delta \textbf{x}(t_{0})$, is given by
\begin{equation} \label{eq:evol}
\delta \textbf{x}(t_0 + T) =  \frac{d\phi_{t_{0}}^{t_{0}+T}(\textbf{x})}{d\textbf{x}} \delta \textbf{x}(t_0) + 
											O(\left\|\delta \textbf{x}\right\|^{2})
\end{equation}
Neglecting the higher order terms, the magnitude of the perturbation is
\begin{equation}
\left\|\delta \textbf{x}(t_0 + T)\right\| = \sqrt{\left\langle \delta \textbf{x}(t_0), \frac{d\phi_{t_{0}}^{t_{0}+T}(\textbf{x})}{d\textbf{x}}^{*} 					
																	\frac{d\phi_{t_{0}}^{t_{0}+T}(\textbf{x})}{d\textbf{x}} \delta \textbf{x}(t_0) \right\rangle}
\end{equation}
The matrix 
\begin{equation}
C = \frac{d\phi_{t_{0}}^{t_0 + T}(\textbf{x})}{d\textbf{x}}^{*} 					
																	\frac{d\phi_{t_{0}}^{t_{0}+T}(\textbf{x})}{d\textbf{x}}
\end{equation}
is the right Cauchy Green deformation tensor. Maximum stretching occurs when the perturbation $\delta  \textbf{x}$ is along the eigenvector $\textbf{n}_{max}$
corresponding to the maximum eigenvalue $\lambda_{max}$ of $C$.  
The growth ratio is given by
\begin{equation}
\left\|\delta \textbf{x}(t_0 + T)\right\| / \left\|\delta \textbf{x}(t_0) \right\| = \textbf{e}^{\sigma_1 (\textbf{x}(t_0)) \left|T\right|} 
\end{equation}
where
\begin{equation}
\sigma_1 (\textbf{x}(t_0)) = \frac{1}{|T|} \ln\sqrt{\lambda_{max}(C)}
\end{equation}
is the maximal finite time Lyapunov exponent.
One can associate an entire spectrum of finite time Lyapunov exponents with $\textbf{x}(t_0)$, ordering them as 
\begin{equation}
\sigma_{1}(\textbf{x}(t_0)) > \sigma_{2}(\textbf{x}(t_0)) > \sigma_{3}(\textbf{x}(t_0)) > \sigma_{4}(\textbf{x}(t_0))
\end{equation}
The entire spectrum of the Lyapunov exponents can be computed from the state transition matrix using singular value decomposition.
\begin{equation}
\Phi(t,t_0) = B(t,t_0) \Lambda (t,t_0)^{1/2} R(t,t_0)
\end{equation}
The diagonal matrix $\Lambda$ gives all the Lyapunov exponents.
\begin{equation}
\Sigma(t_f,t_0) = \ln [\Lambda(t_f,t_0)^{1/2|T|}]
\end{equation}
where $T= t_f - t_0$ and $\Sigma(t_f,t_0) = diag(\sigma_1,...,\sigma_4)$.
An arbitrary perturbation in the fixed basis can be transformed using a time dependent transformation \cite{WigginsIde}.
\begin{equation}
\delta \textbf{x}'(t) = A(t,t_0,t_f)\delta \textbf{x}(t)
\end{equation}
such that in the new basis (the primed frame), the variational equations become
\begin{equation}
\delta \dot{{\textbf{x}}}'(t)= \Sigma(t_f,t_0)\delta \textbf{x}'(t)
\end{equation}
Since $\Sigma(t_f,t_0)$ is a constant diagonal matrix, we have

\begin{equation}
\delta {\textbf{x}}'(t)= e^{(t-t_0)\Sigma(t_f,t_0)}\delta \textbf{x}'(t_0)
\end{equation}

The first coordinate in the new frame grows as $ \delta x'_1(t) = e^{(t-t_0) \sigma_1} \delta x'_1(t_0)$.
The time dependent transformation $A(t)$ is given by \cite{WigginsIde},
\begin{equation}
A(t,t_0,t_f) = e^{(t-t_0)\Sigma(t_f,t_0)}R(t_f,t_0)^* R(t,t_0)\Sigma(t,t_0)^{-1/2}B(t,t_0)
\end{equation}

\subsection{Sensitivity to initial relative velocity}
Since the dynamics of the inertial particle is in a four-dimensional phase space, the separatrices, that is LCS defined by ridges in the field of the maximal FTLE, are three dimensional surfaces (see \cite{Shadden2}).
However, because the system is dissipative and the global attractor is the $xy$ subspace, we can obtain meaningful information by restricting the computations to a lower dimensional subdomain of the phase space. This we do by considering an initial perturbation only in the relative velocity subspace and study how this perturbation grows in the $xy$ plane, the configuration space, i.e,
\begin{equation} \label{perturb}
\delta \textbf{x}(t_0) = [0,0,\Delta W_x, \Delta W_y]^{*}
\end{equation}
where $\Delta W_x, \Delta W_y$ are the perturbations in the relative velocity subspace.
Using the time dependent transformation $A(t,t_0,t_f )$ the evolution of the perturbation is given by
\begin{equation}
\delta {\textbf{x}}(t)= A^{-1}(t)e^{(t-t_0)\Sigma(t_f,t_0)} A(t_0) \delta \textbf{x}(t_0).
\end{equation}

The growth of perturbation in the $xy$ plane is given by the first two components of the above vector.
The last two components of the above vector are the evolution of the perturbations in the relative velocity subspace. Since the $xy$ plane is a global attractor these tend to zero. One can choose a finite time, $T$, such that the evolution of the initial perturbation comes arbitrarily close to the $xy$ plane. In this way the sensitivity of the final spatial location of the particles to initial relative velocity can be computed.

\subsection{Numerical computation of the Sensitvity Field}
As equation(15) shows the evolution of a perturbation is along the four basis vectors. For an arbitrarily oriented initial perturbation the growth may not be dominated in the direction of greatest expansion for short integration times. This can be overcome by sampling multiple perturbations in the different directions. A reference point and its neighbors are identified and after a finite time their positions in configuration space are computed. The state transition matrix can then be computed at each point in the $xy$ plane, by using a central finite difference method. For initial perturbations restricted to $W_x W_y$ subspace, this gives

\begin{equation} \label{finite_diff}
\phi_{,rW} = 
\left(\begin{array}{cc}
	\frac{x_{i,j,k+1,l}(t_0 + T)-x_{i,j,k-1,l}(t_0 + T)}{\Delta W_x(t_0)} & \frac{x_{i,j,k,l+1}(t_0 + T)-x_{i,j,k,l-1}(t_0 + T)}{\Delta W_y(t_0)}\\
	\frac{y_{i,j,k+1,l}(t_0 + T)-y_{i,j,k-1,l}(t_0 + T)}{\Delta W_x(t_0)} &  \frac{y_{i,j,k,l+1}(t_0 + T)-y_{i,j,k,l-1}(t_0 + T)}{\Delta W_y(t_0)}
\end{array}\right)
\end{equation}
The relative velocity sensitivity field , $\sigma (W_x,W_y)$ is given by,
\begin{equation}
\sigma (W_x,W_y)= \frac{1}{|T|}\ln\sqrt{\lambda_{max}(\phi_{,rW} ^{*} \phi_{,rW})}  
\end{equation}
Ridges on this sensitivity surface are one dimensional structures similar to LCS. The ridges in the maximal sensitivity field $\sigma (W_x,W_y)$, partition the relative velocity subspace. We applied the above procedure to a cellular flow.
\\ \textit{A note on the terminology} - The field measuring the sensitivity of the final location of particles in configuration space with respect to perturbations in initial relative velocity is analogous to the FTLE field, but not identical. To obtain the true FTLE field, one would have to compute the $4 \times 4$ state transition matrix, $\Phi$. Using the notation of eq \eqref{finite_diff},

\begin{equation} 
\Phi =  
\left(\begin{array}{cc}
	\phi_{,rr} & \phi_{,rW}\\
	\phi_{,Wr} & \phi_{,WW}
\end{array}\right)
\end{equation}
The FTLE field is then given by $\sigma(x,y,W_x,W_y)= \frac{1}{|T|}\ln\sqrt{\lambda_{max}(\Phi ^{*} \Phi)}$. Ridges in this field are 3 dimensional structures and represent the true LCS.

\section{Example: cellular flow }
This flow is described by the stream function 
\begin{equation}\label{ti_flow}
\psi (x,y,t) = a\cos{x} \cos{y}
\end{equation}
The velocity field is given by,

\begin{equation}
u = -a \cos{x} \sin{y}
\end{equation}

\begin{equation}
v = a \sin{x} \cos{y}
\end{equation}

\begin{figure}
\begin{center}
\includegraphics[height=0.5 \textwidth]{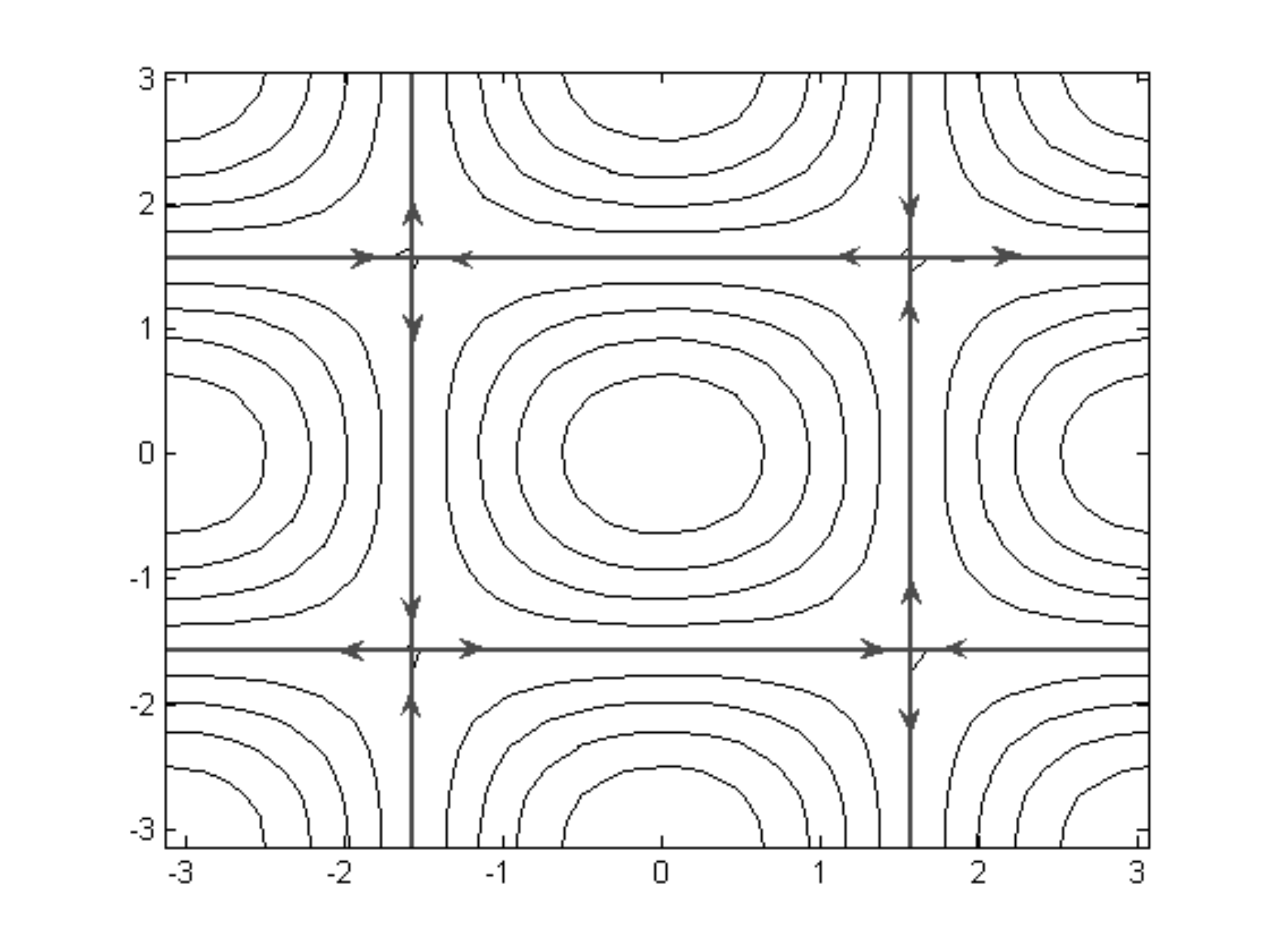}
\caption{\label{streamlines}{\footnotesize{Streamlines of $ \psi$ $= a $ $\cos x \cos y$ form an array of cells. The arrows indicate the heteroclinic trajectories connecting the fixed points of the velocity field formed by $ \psi $. For this velocity field, the heteroclinic trajectories coincide with the LCS}}}
\end{center}
\end{figure}

\begin{figure}[h,t]
\begin{center}
\begin{tabular}{ccc}
\includegraphics[height=0.23 \textwidth]{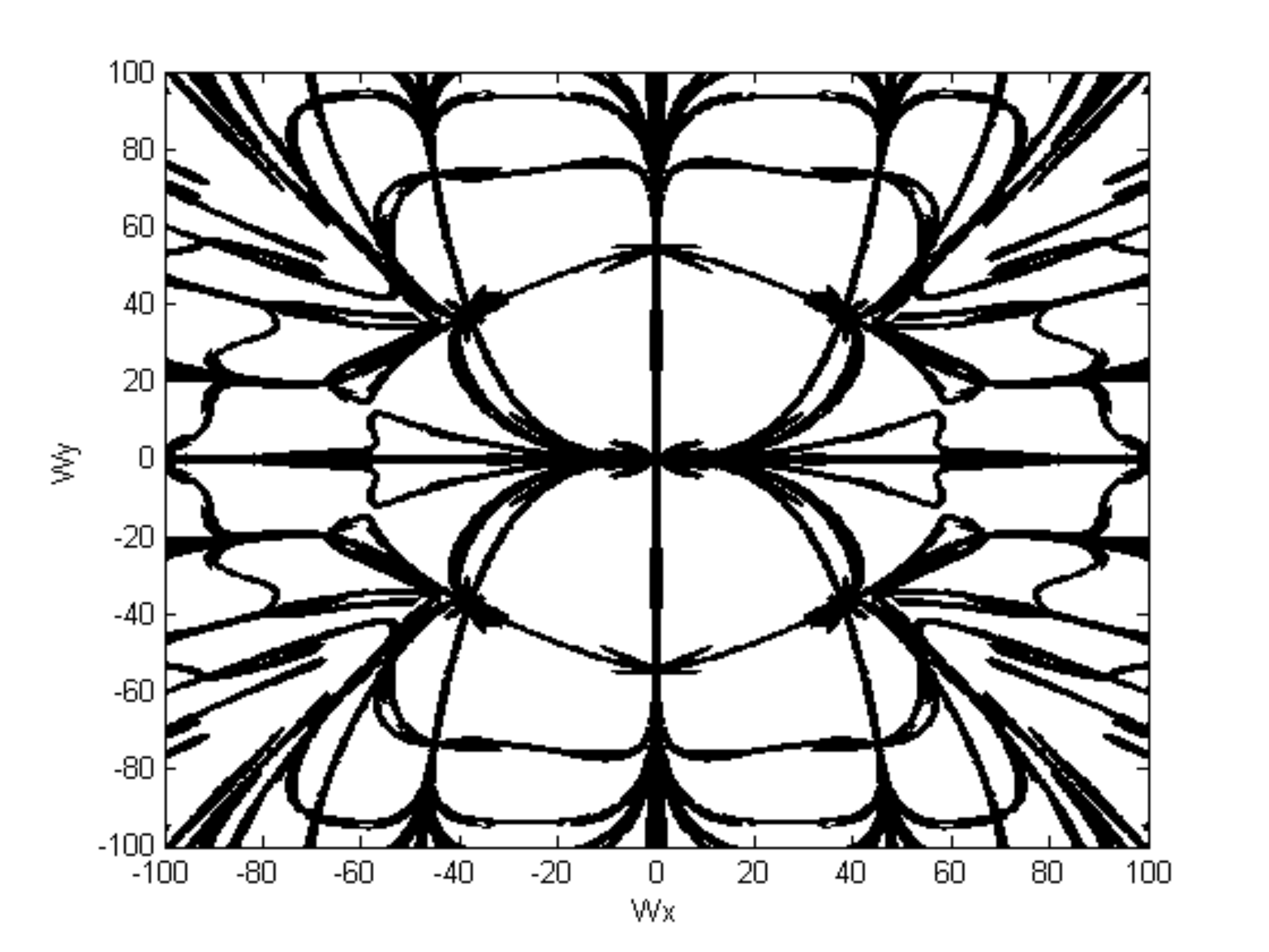} &
\includegraphics[height=0.23 \textwidth]{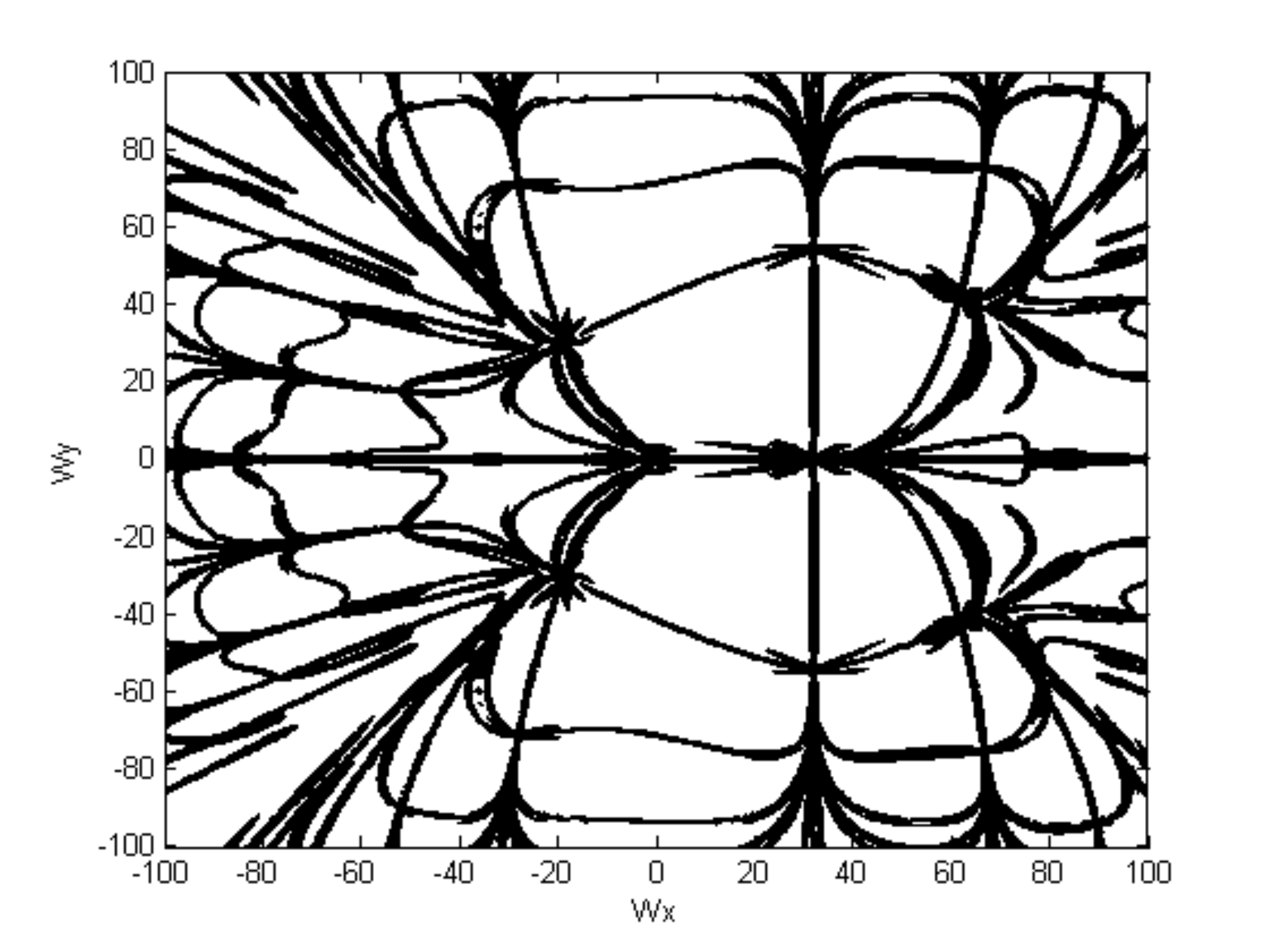} & 
\includegraphics[height=0.23 \textwidth]{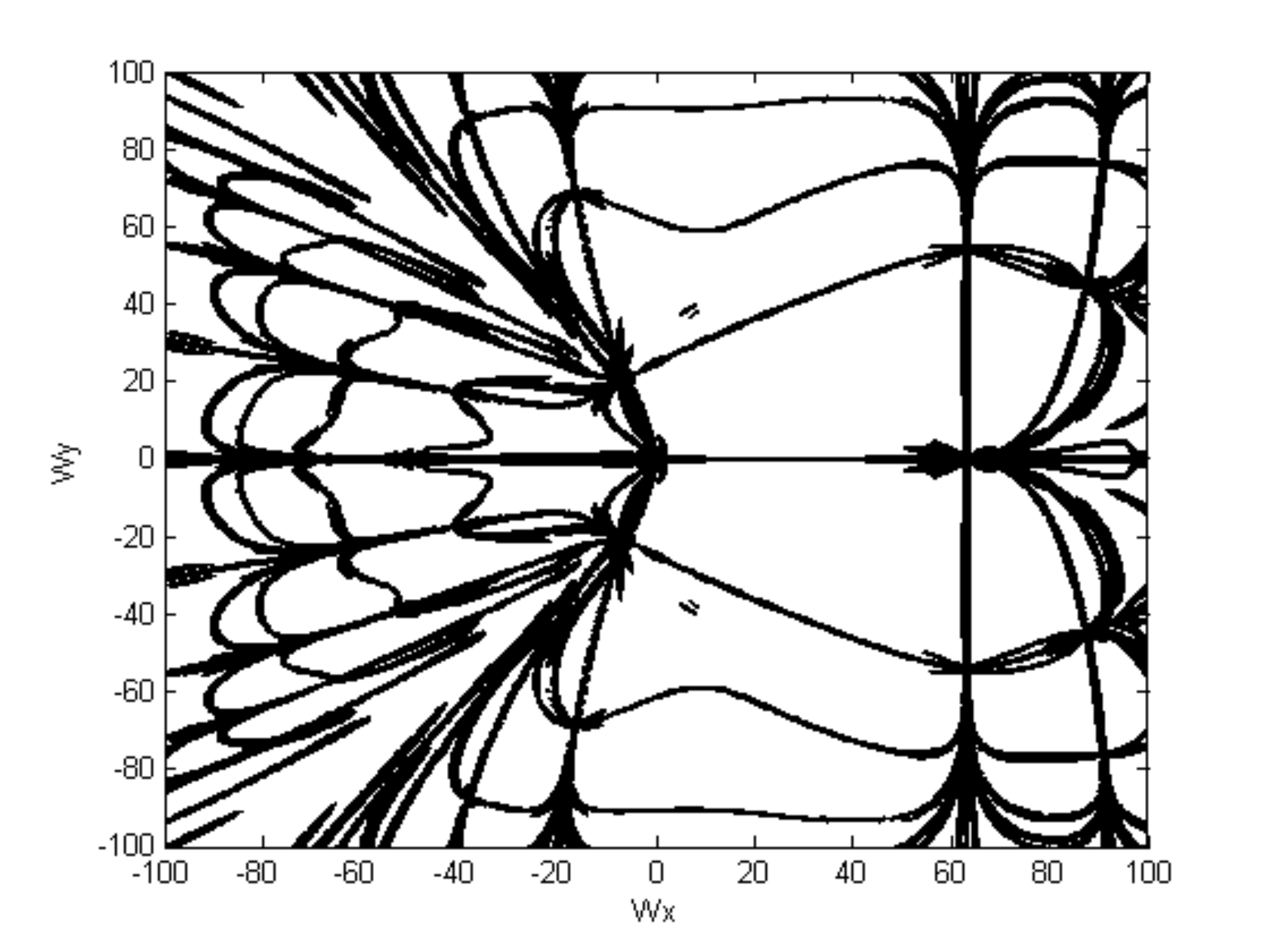} \\
{\footnotesize (a)} & {\footnotesize (b)}& {\footnotesize (c)}
\end{tabular}

\begin{tabular}{ccc}
\includegraphics[height=0.23 \textwidth]{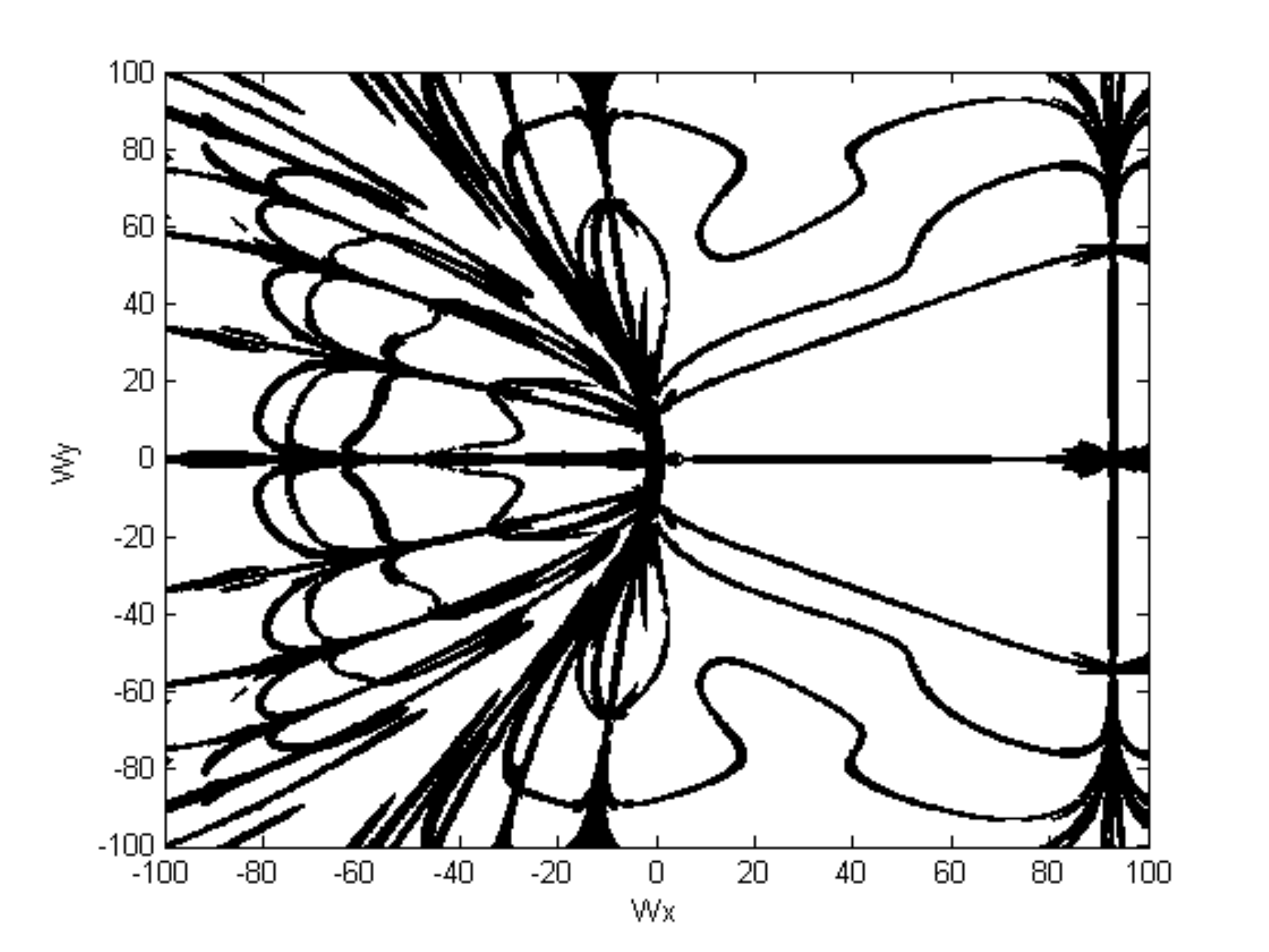} &
\includegraphics[height=0.23 \textwidth]{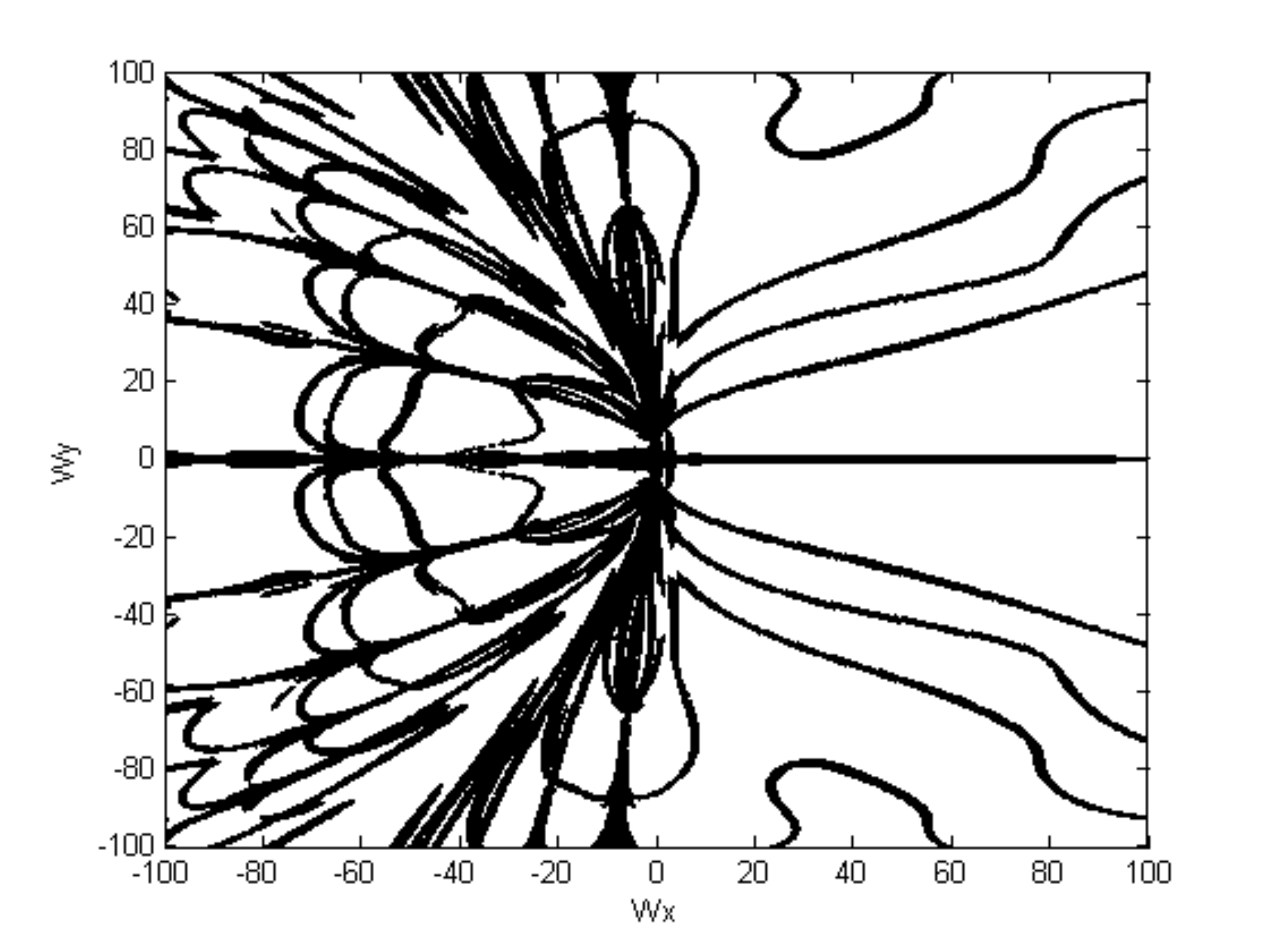} & 
\includegraphics[height=0.23 \textwidth]{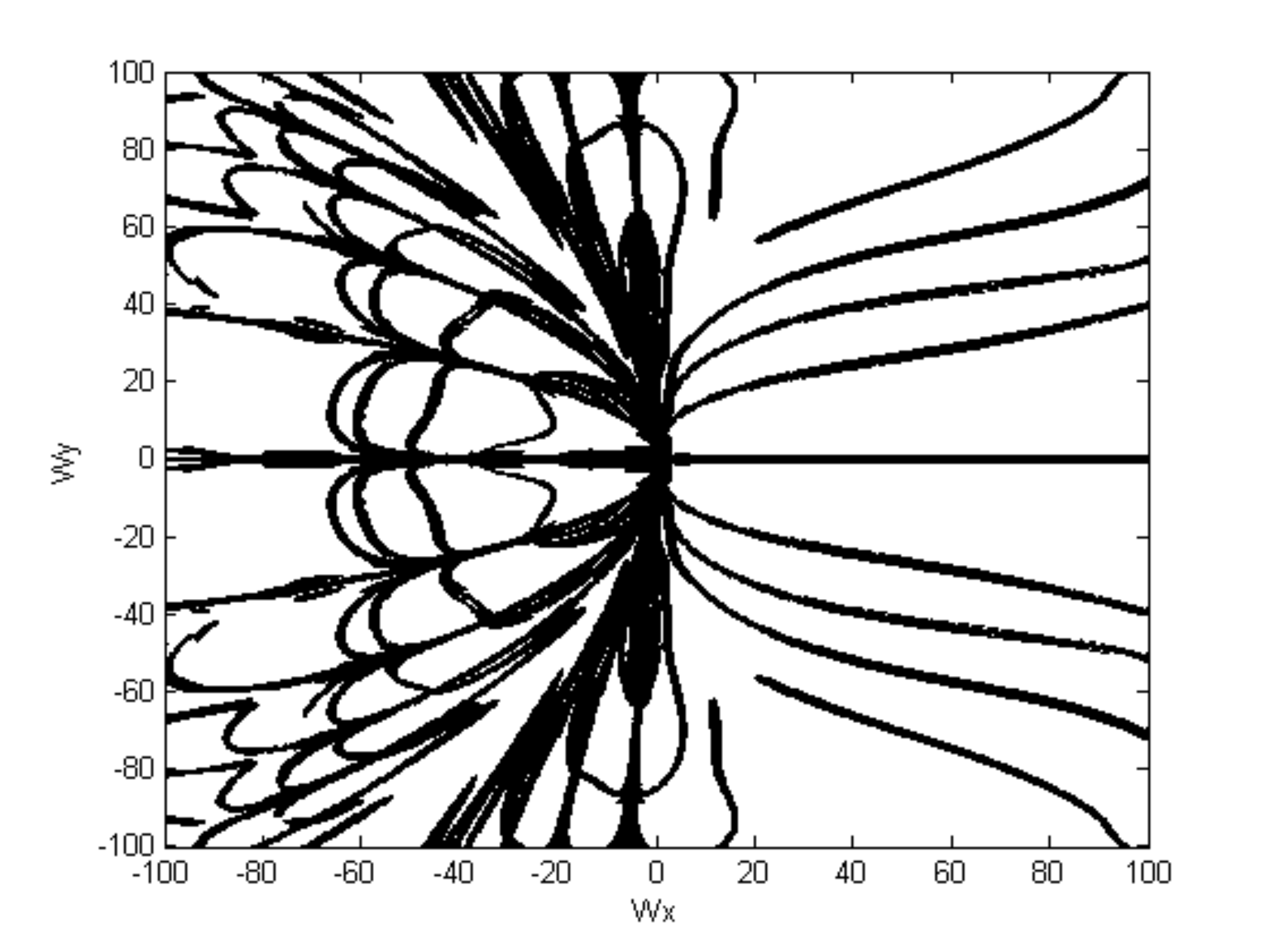} \\
{\footnotesize (d)} & {\footnotesize (e)} & {\footnotesize (f)}
\end{tabular}

\begin{tabular}{ccc}
\includegraphics[height=0.23 \textwidth]{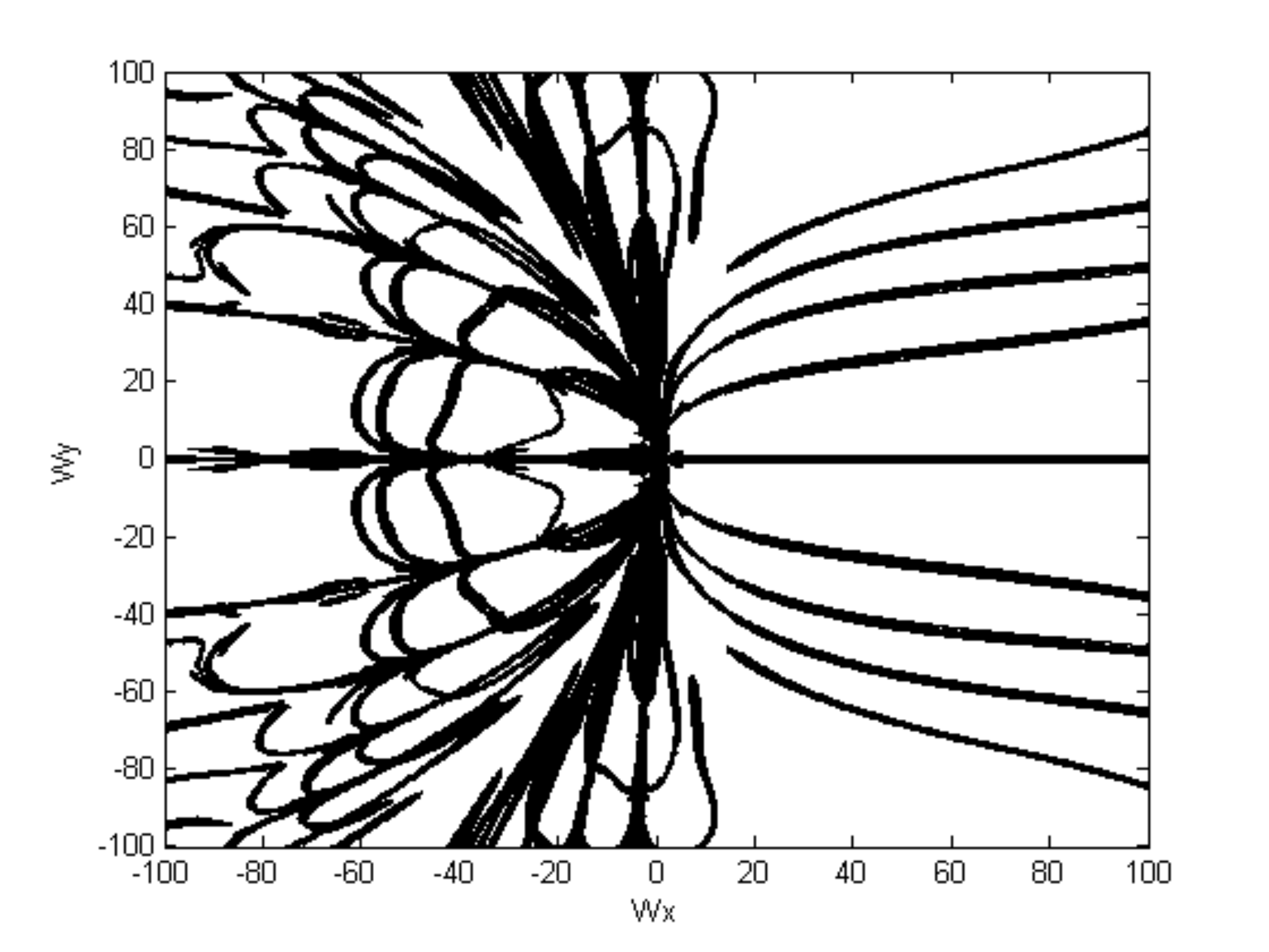} &
\includegraphics[height=0.23 \textwidth]{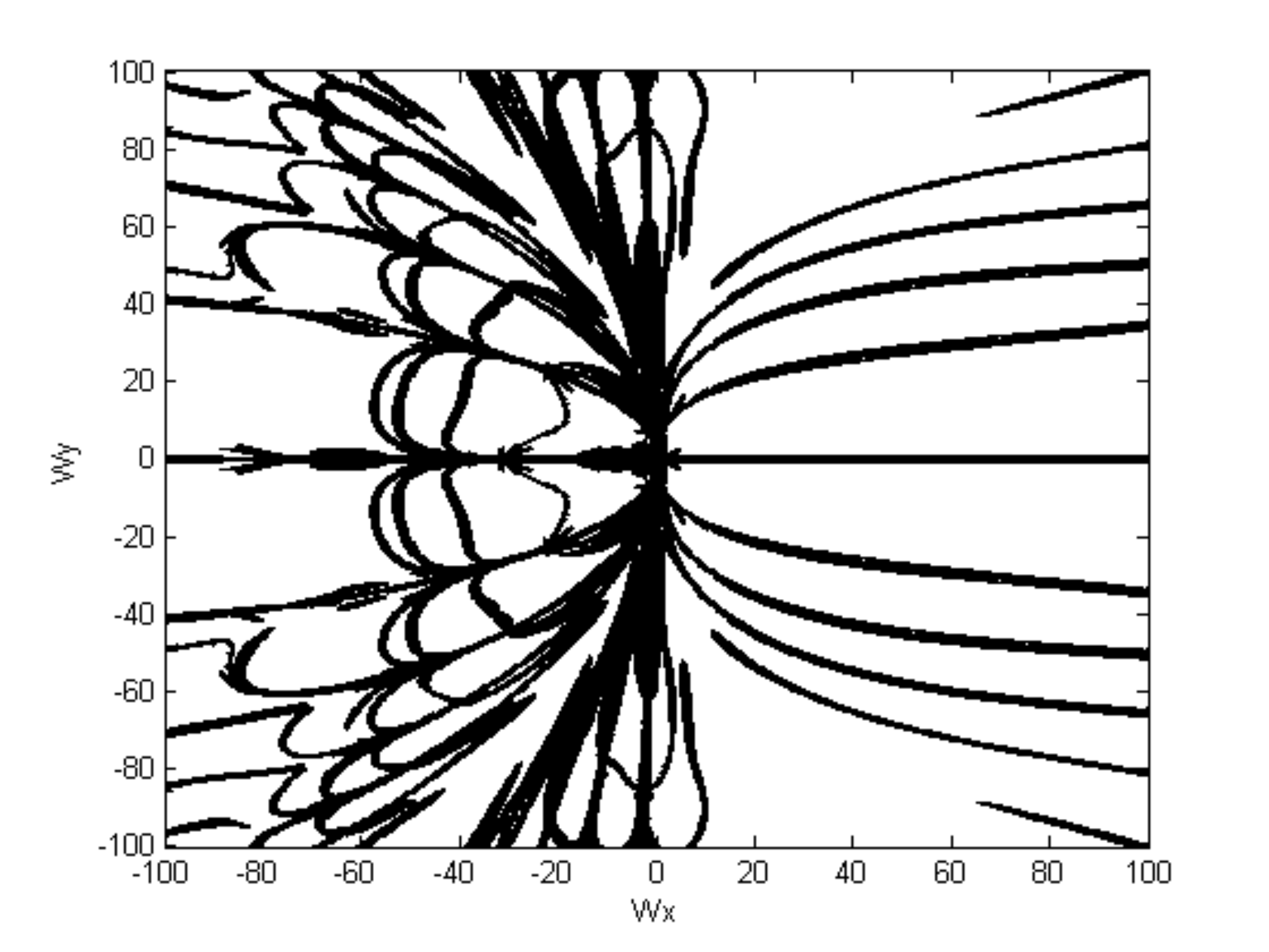} &
\includegraphics[height=0.23 \textwidth]{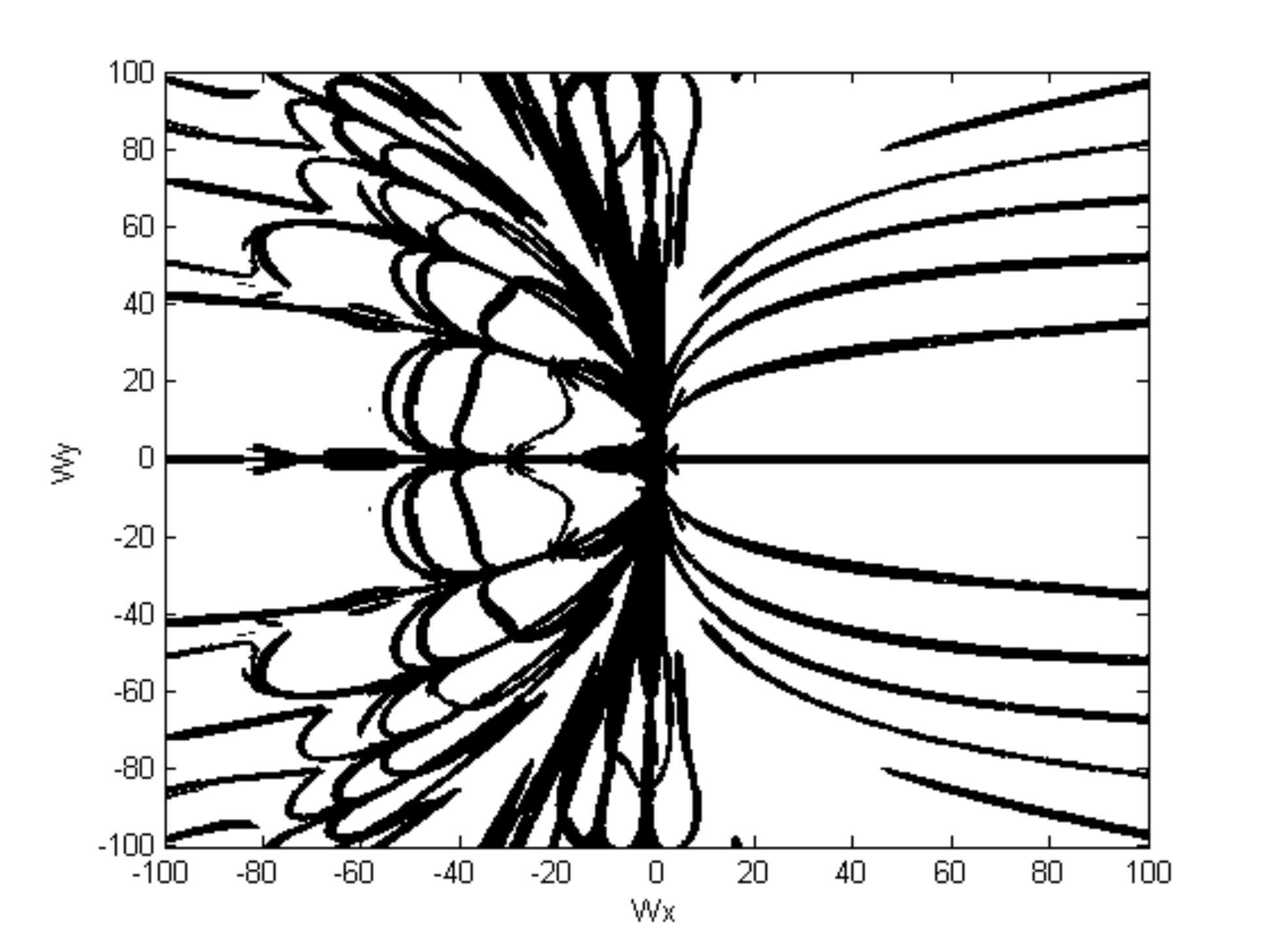} \\
{\footnotesize (g)} & {\footnotesize (h)} & {\footnotesize (i)}
\end{tabular}
\newpage

\begin{tabular}{ccc}
\includegraphics[height=0.23 \textwidth]{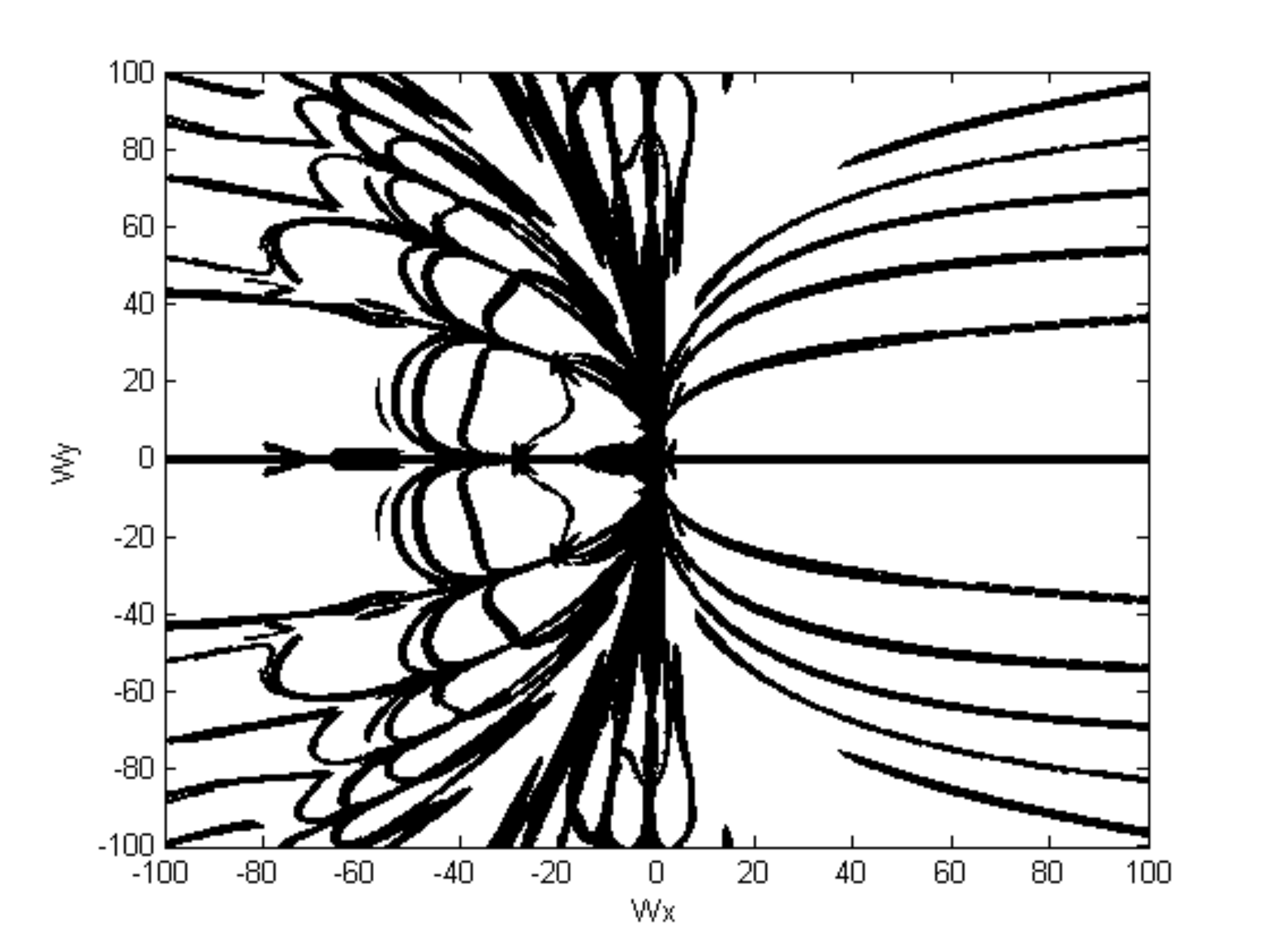} & 
\includegraphics[height=0.23 \textwidth]{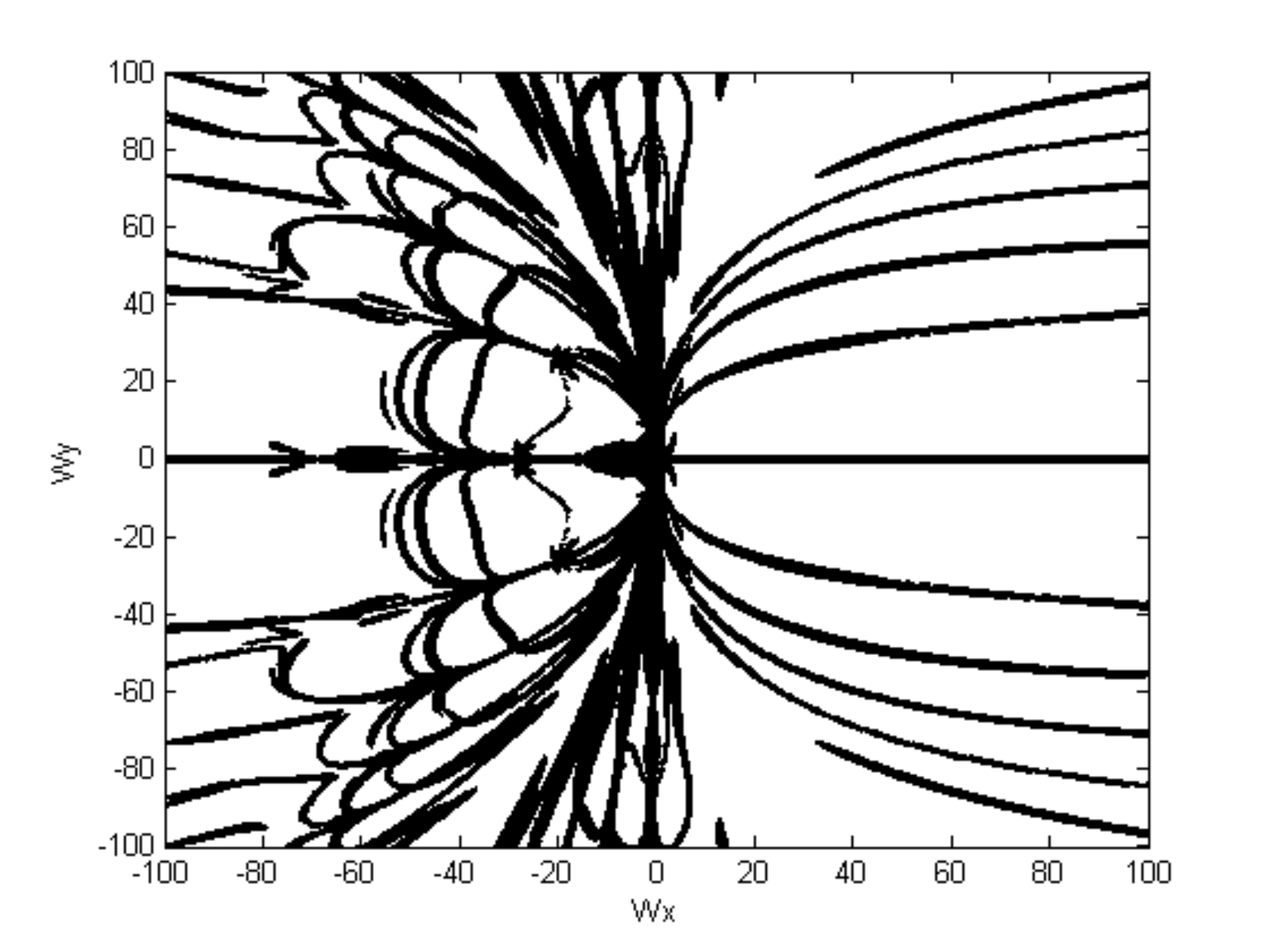}  &
\includegraphics[height=0.23 \textwidth]{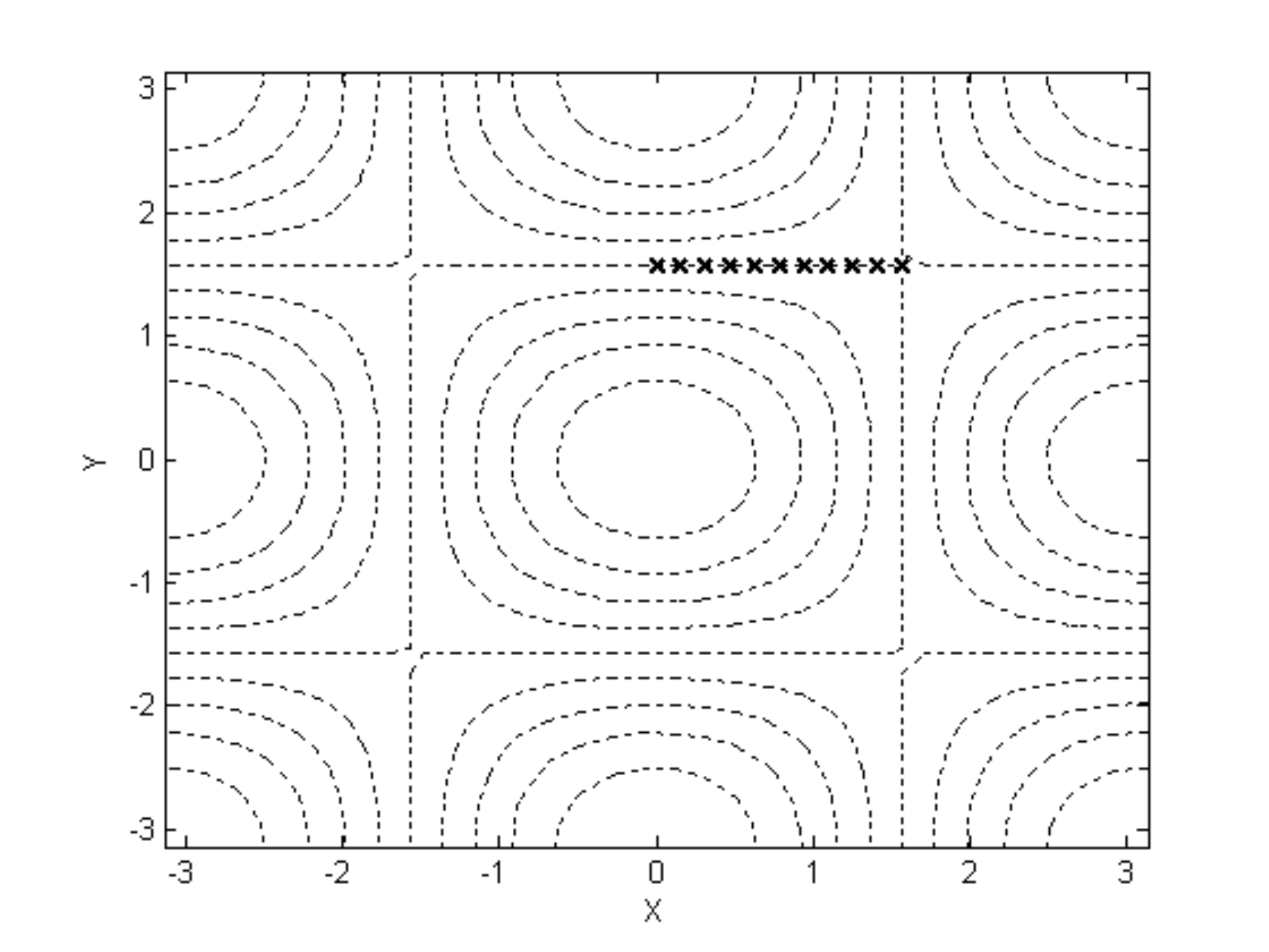}  \\
 {\footnotesize (j)} & {\footnotesize (k)} & {\footnotesize (l)}
\end{tabular}

\end{center}
\caption{\label{variation of FTLE}{\footnotesize{Ridges in the sensitivity field for $ \psi$ $= a $ $\cos x \cos y$. Initial spatial position varies from $(x_0,y_0) = (\pi/2, \pi/2)$ to $(0, \pi/2)$, at the points shown in (l),along $\psi = 0$, at intervals of $0.05\pi$. The plots show a smooth variation in the structure of the ridges in the sensitivity field. Parameters , a = 100, St = 0.2, integration time T = 0.24}}}
\end{figure}

There are heteroclinic connections from the stable and unstable manifolds of the fixed points $(2n+1)\frac{\pi}{2}$, shown by the arrows in Figure \ref{streamlines}, which are also the boundaries of the cells. These coincide with LCS$_f$, the LCS for the fluid velocity field. The LCS$_f$ is to be distinguished from the LCS of the inertial particle in the full four dimensional phase space. By choosing initial perturbations of the form given by eq.~\eqref{perturb} at different points along a streamline, we follow how these perturbations grow in the $xy$ plane by integrating the particle trajectories numerically from which the sensitivity field is computed. Figure \ref{variation of FTLE} shows the sensitivity field computed for initial perturbations in the relative velocity subspace, at different points on the streamline $\psi(x,y,t) = 0$. The ridges in this field have high values of sensitivity. It can be seen that there is a continuous variation in the ridges of the sensitivity field with respect to the initial $(x,y)$ coordinates. In each case the sensitivity field at a given point depends on the underlying LCS$_f$ of the fluid flow.

\begin{figure}[h,t]
\begin{center}
\includegraphics[height =0.33\textwidth]{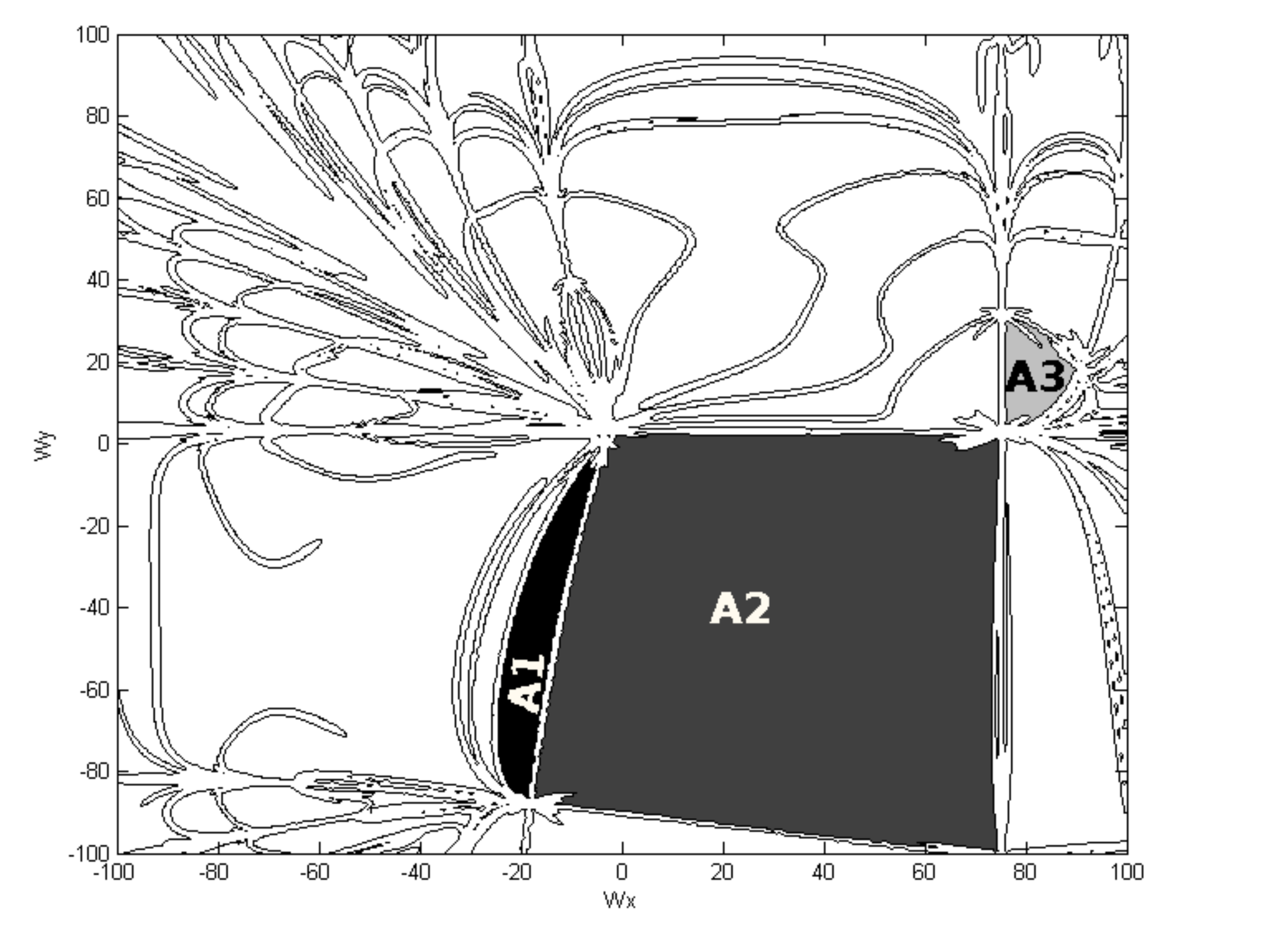}\\
{\footnotesize (a)} \\
\includegraphics[height =0.33\textwidth]{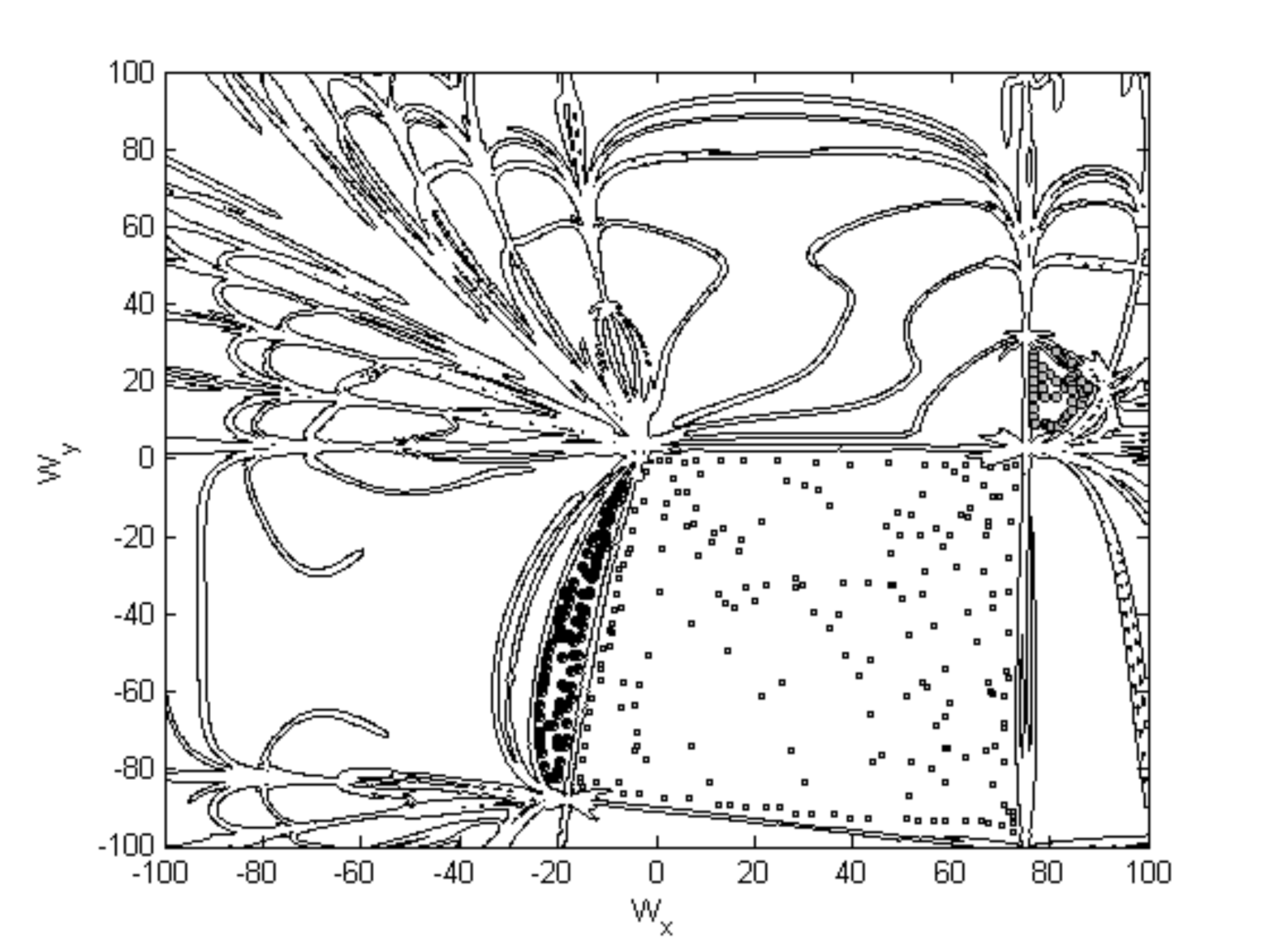}\\
{\footnotesize (b)} \\
\includegraphics[height =0.33\textwidth]{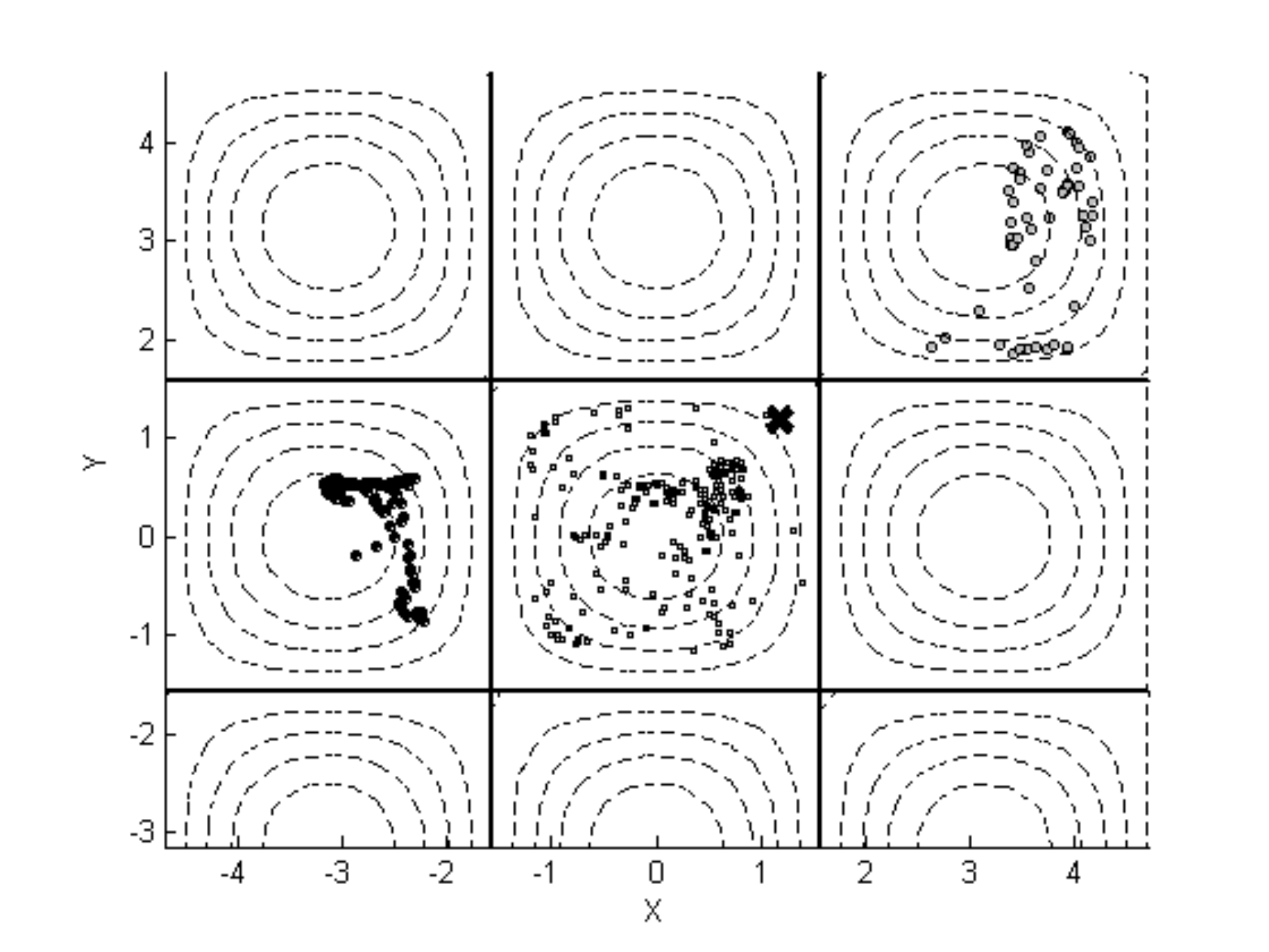}\\
{\footnotesize (c)}
\end{center}\caption{\label{velocity_partition}{\footnotesize(a) Ridges in the sensitivity field, partition the velocity subspace into regions of distinct qualitative dynamics. Three such partitioned regions are shown. 
(b) Particles starting with relative initial velocities belonging to distinct partitions in the relative velocity subspace are segregated into different cells in the $xy$ plane. Time of integration T = 0.24, St = 0.2.  The initial position of all particles is $(x_0,y_0) = (3 \pi /8, 3 \pi /8)$, shown by the $\mathbf{x}$ marker.
}}
\end{figure}

The ridges in the sensitivity field have meaningful information about the dynamics of inertial particles even when computed at points far from the saddle points of the fluid flow. This is shown in Figure \ref{velocity_partition}(a) which is the sensitivity field computed at $(x,y)= (\frac{3\pi}{8},\frac{3\pi}{8})$.
The ridges in the sensitivity field partition the relative velocity subspace according to the final location of particles. In Figure \ref{velocity_partition}(b) the ridges in the partial FTLE field are used to identify regions in the relative velocity subspace, that produce qualitatively different trajectories. 
Particles that start at the same physical location, but are in different regions of the relative velocity subspace, are neatly separated from particles that started in other regions, as shown in Figure \ref{velocity_partition}(c). Thus the ridges in the sensitivity field, have the property of a separatrix.

\section{Segregation of particles by Stokes number }

\begin{figure}[h,t]
\begin{center}
\includegraphics[height=0.32\textwidth]{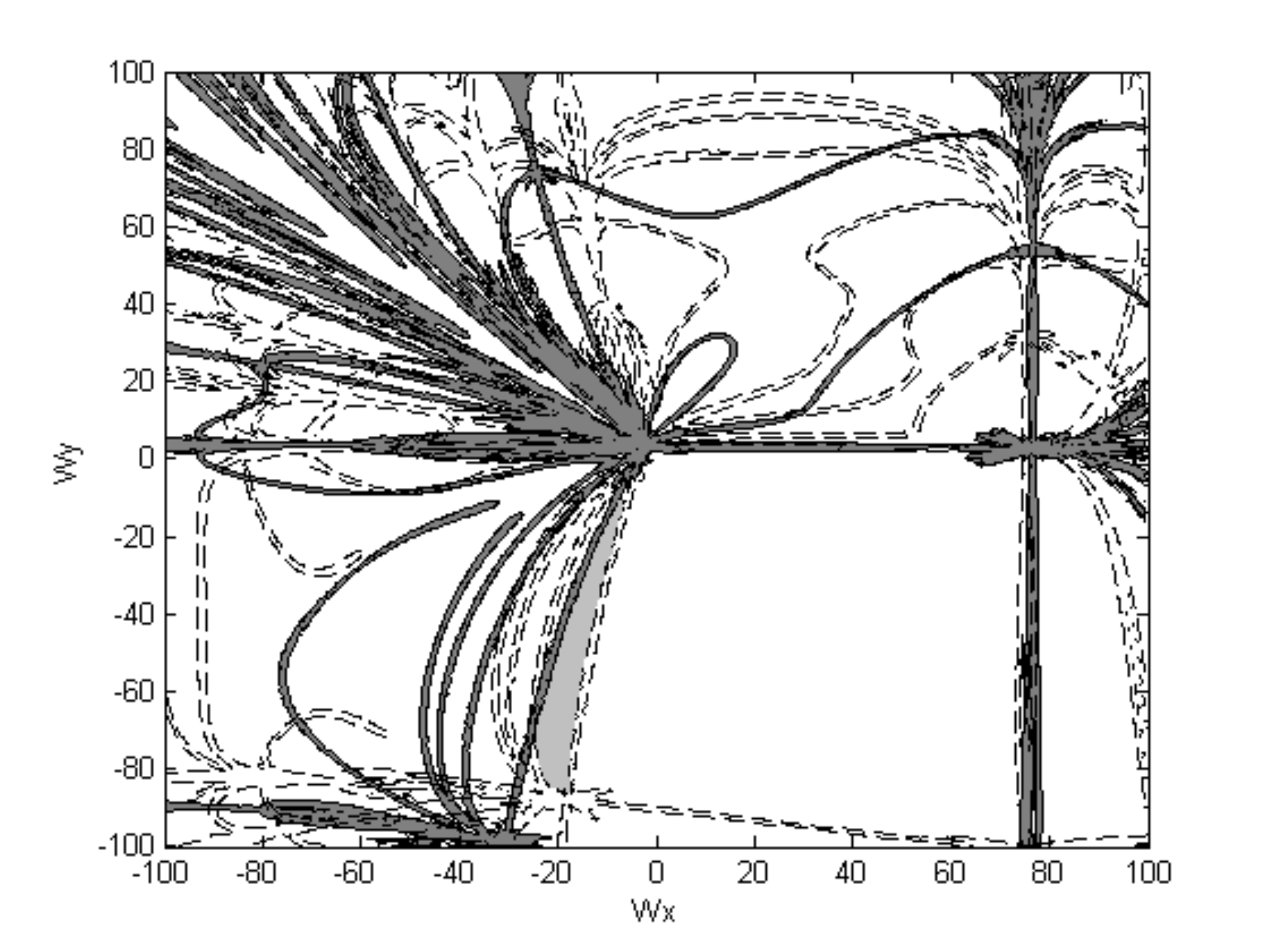}\\
{\footnotesize (a)}\\

\begin{tabular}{ccc}
\includegraphics[height=0.21 \textwidth]{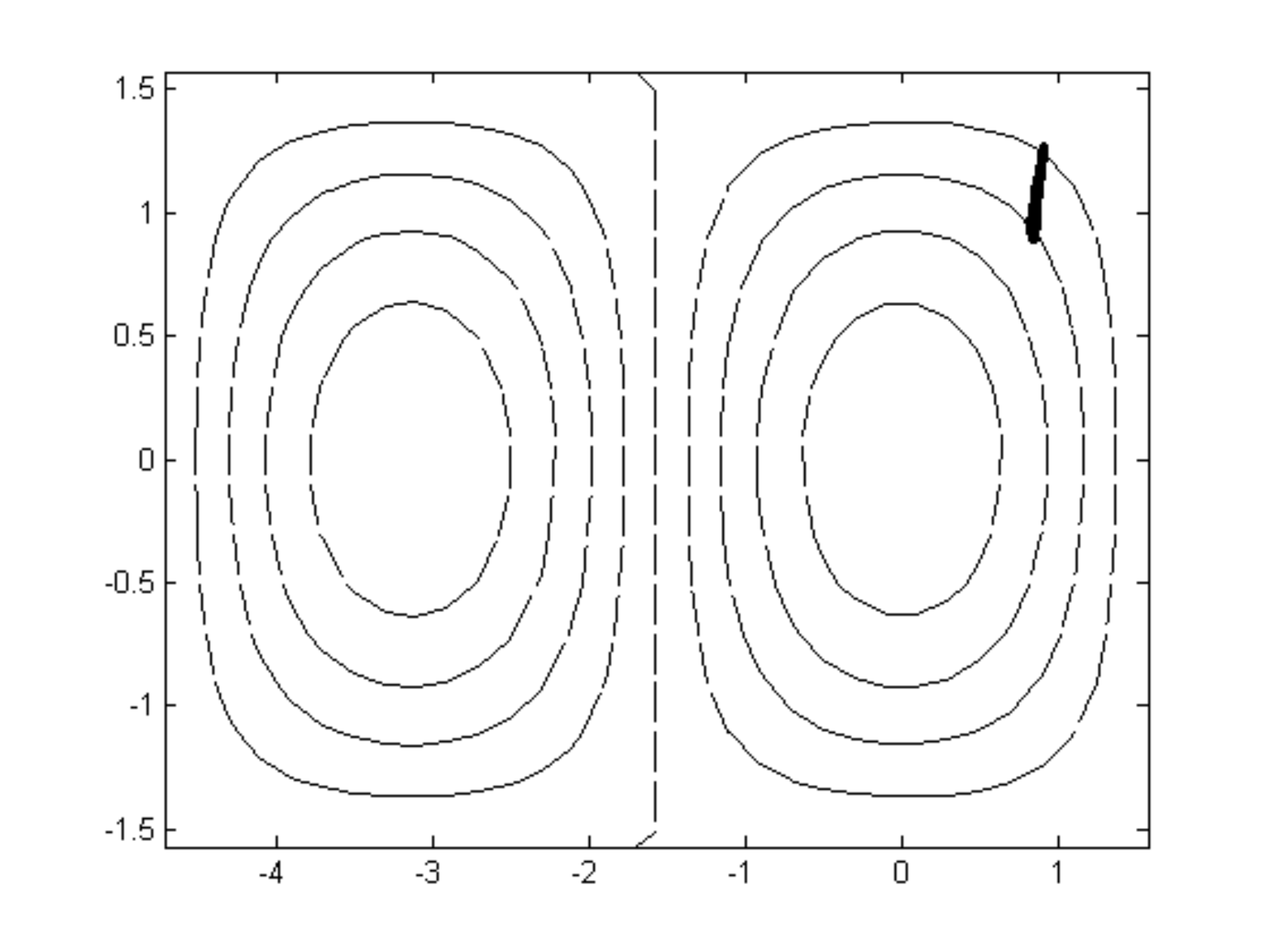} &
\includegraphics[height=0.21 \textwidth]{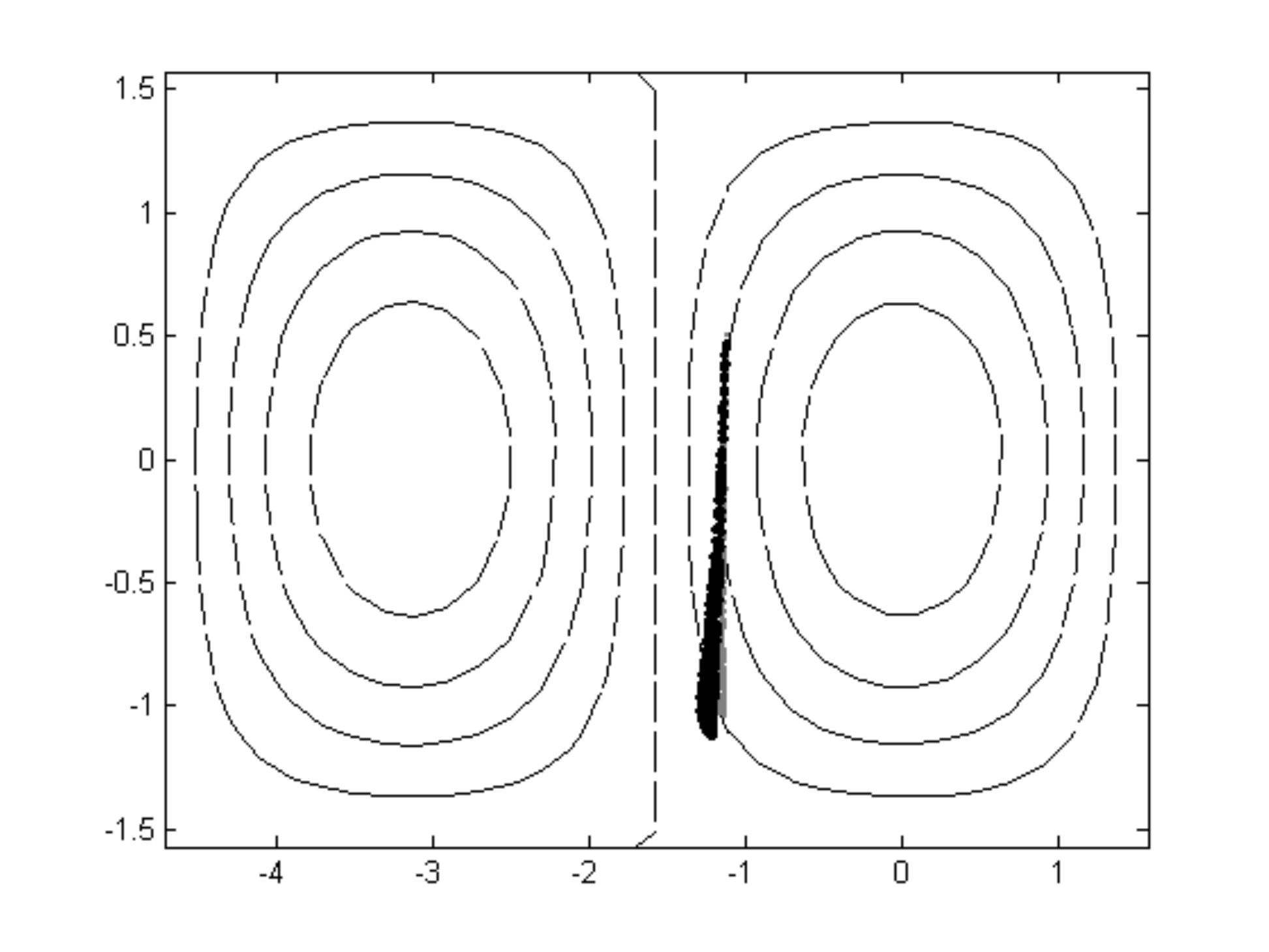} &
\includegraphics[height=0.21 \textwidth]{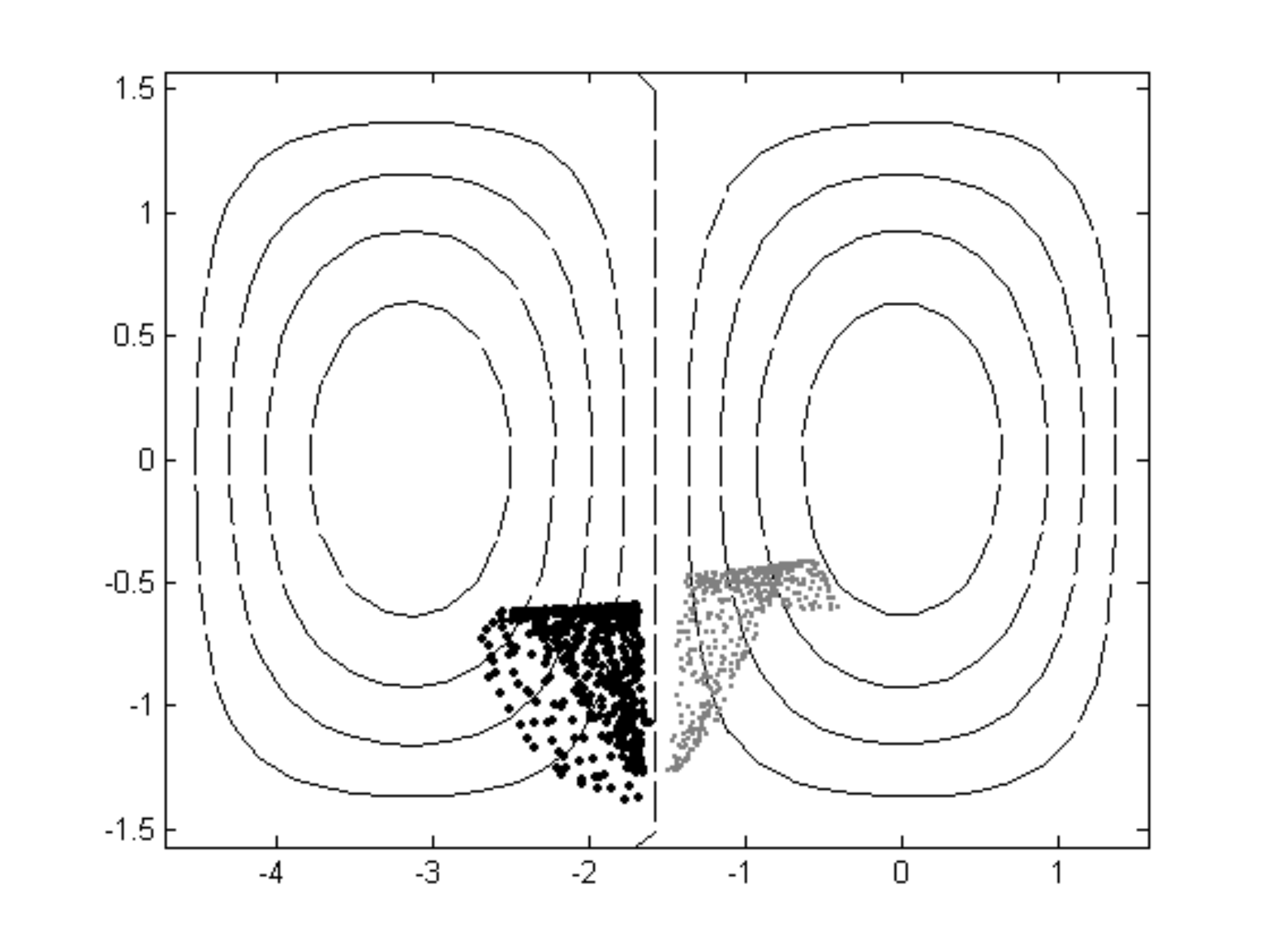} \\
{\footnotesize (b)T = 0.005} & {\footnotesize (c)T = 0.030} & {\footnotesize (d)T = 0.060}
\end{tabular}

\begin{tabular}{ccc}
\includegraphics[height=0.21 \textwidth]{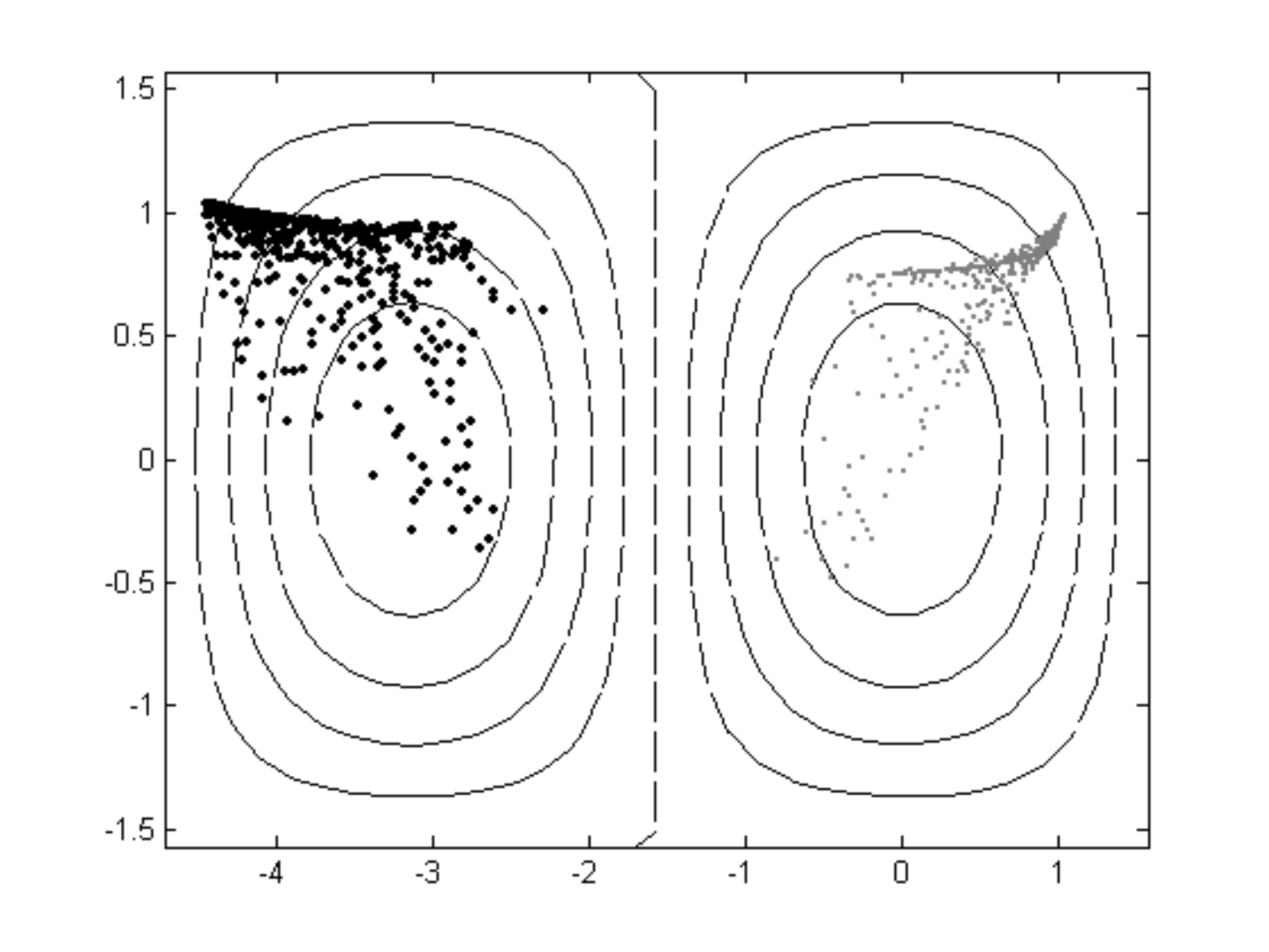} &
\includegraphics[height=0.21 \textwidth]{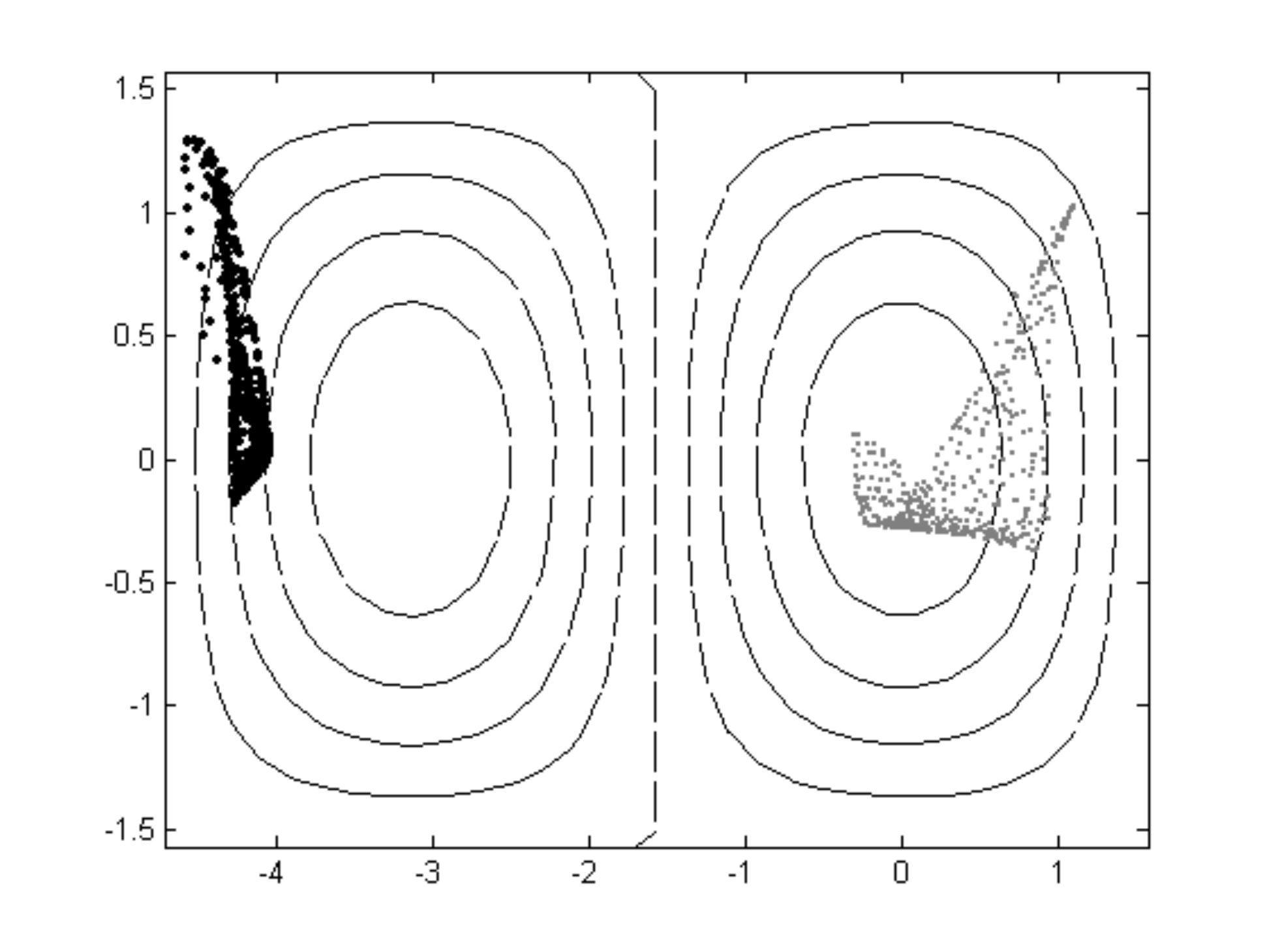} &
\includegraphics[height=0.21 \textwidth]{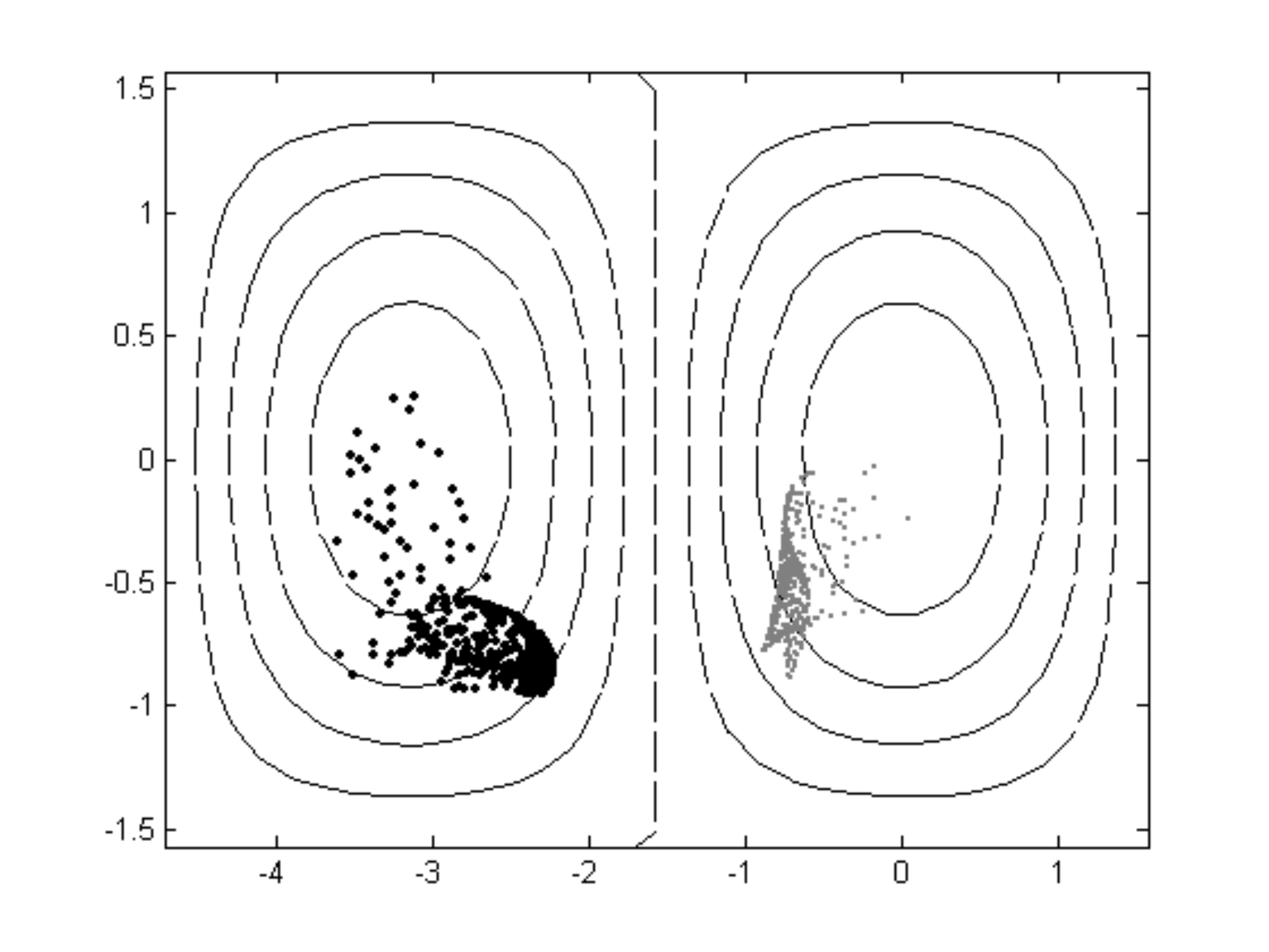} \\
{\footnotesize (e)T = 0.085} & {\footnotesize (f)T = 0.110} & {\footnotesize (g)T = 0.135}
\end{tabular}

\begin{tabular}{ccc}
\includegraphics[height=0.21 \textwidth]{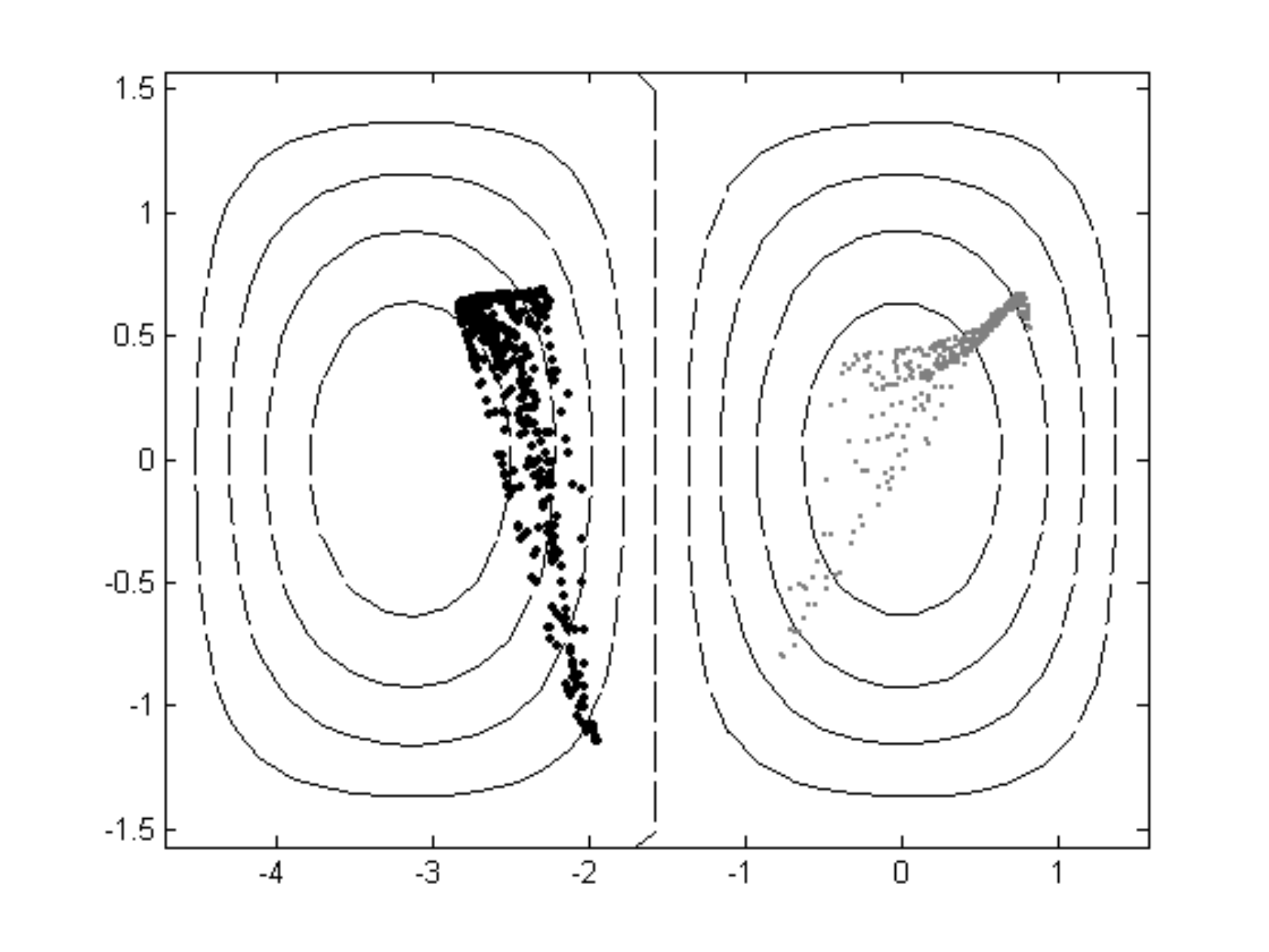} &
\includegraphics[height=0.21 \textwidth]{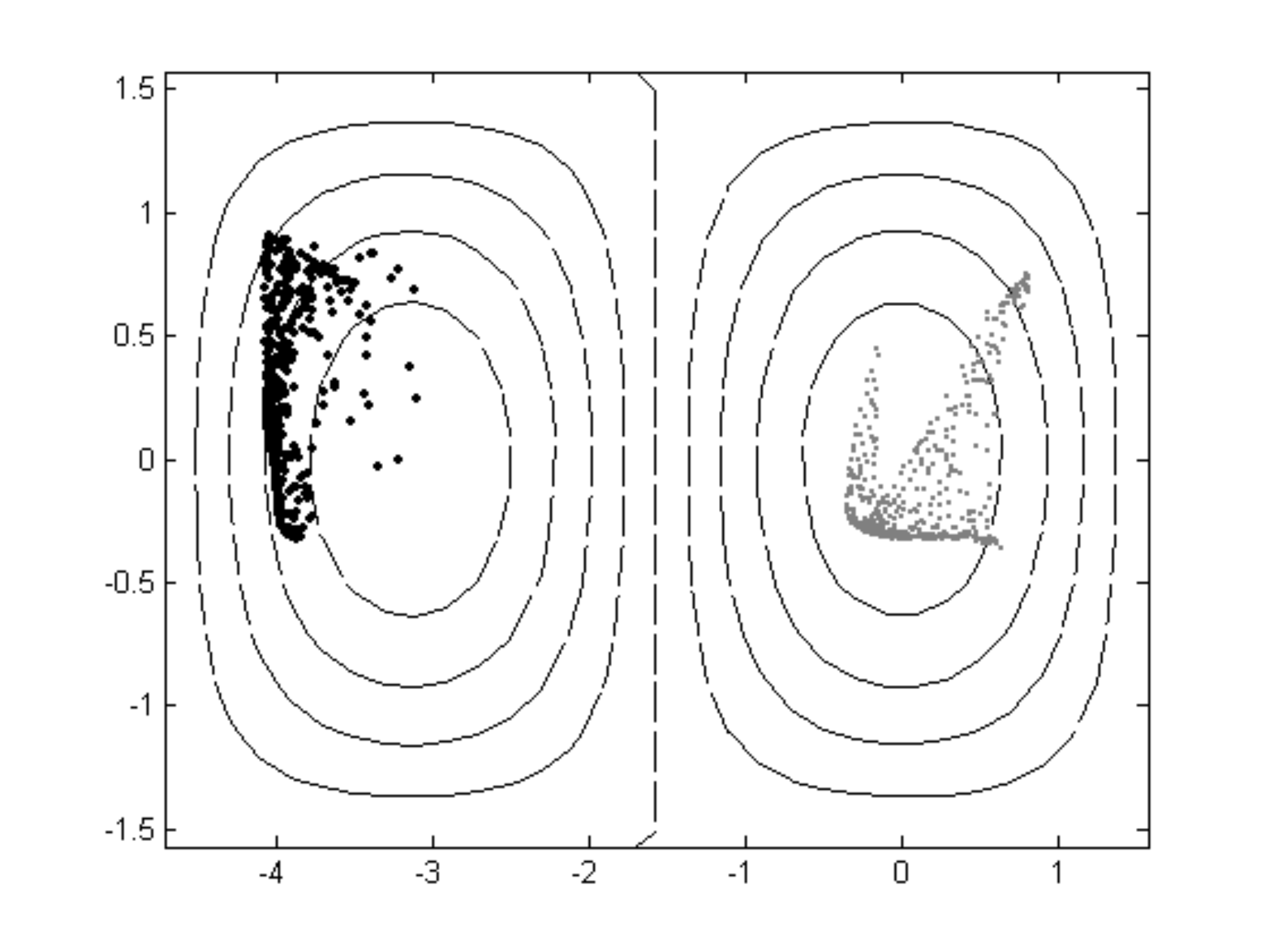} &
\includegraphics[height=0.21 \textwidth]{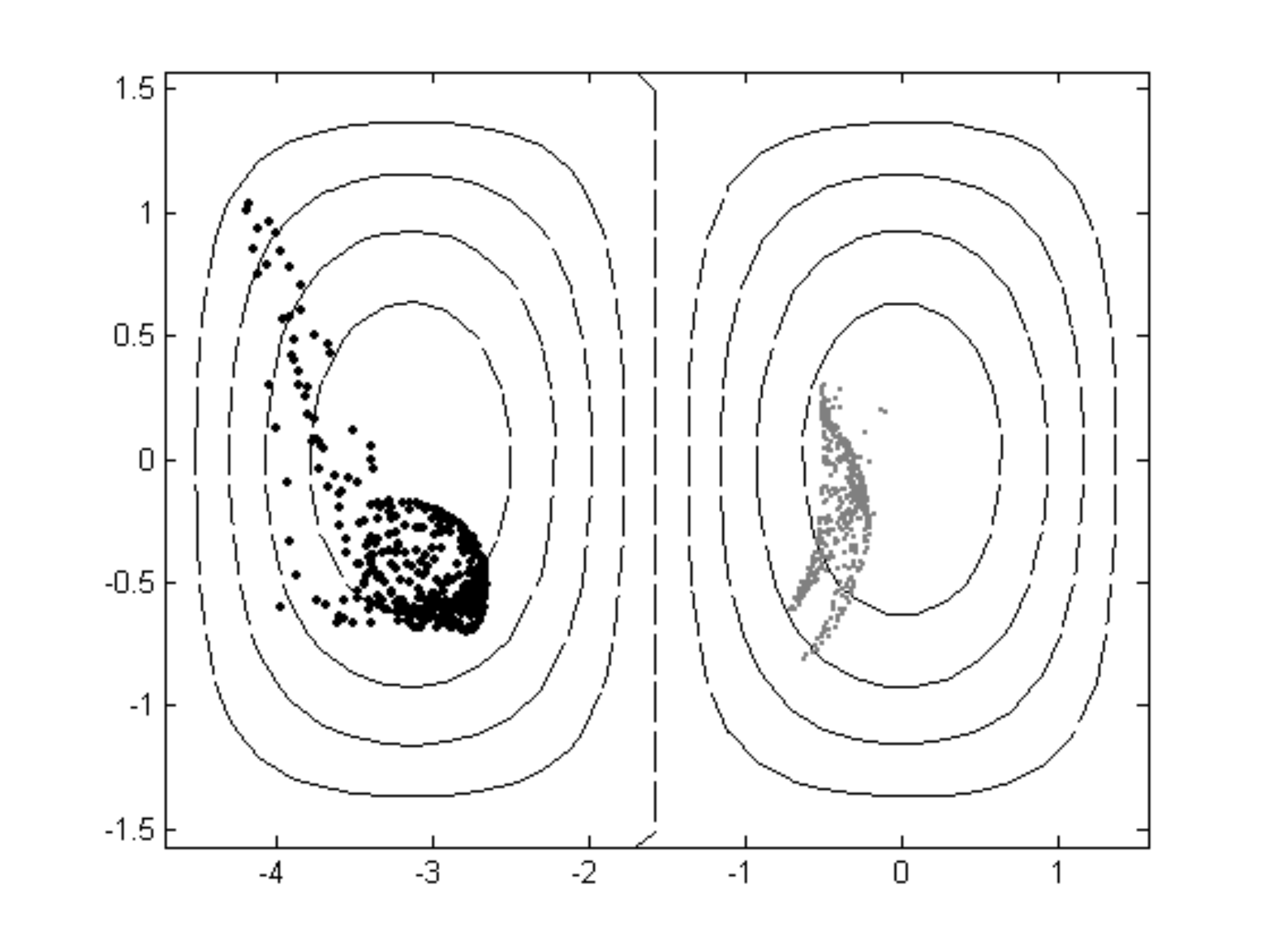} \\
{\footnotesize (h)T = 0.160} & {\footnotesize (i)T = 0.185} & {\footnotesize (j)T = 0.210}
\end{tabular}

\end{center}\caption{\label{particle_separation}{\footnotesize
(a) Ridges in the sensitivity field for particles with St =0.2 (hatched) and St=0.1(thick) respectively. The initial position of all particles is $(x_0,y_0) = (3 \pi /8, 3 \pi /8)$. The grey patch is a sample region sandwiched between the ridges of the two Stokes numbers. 
(b) Particles starting at $(x_0,y_0) = (3 \pi /8, 3 \pi /8)$ and relative initial velocity belonging to the grey patch, are separated into different cells in the $xy$ plane. 
}}
\end{figure}

Eq.~\eqref{eq:dissipative1} can be diagonalised as 
\begin{equation}
\frac{d\textbf{W}_{d}}{dt} = 
\left(\begin{array}{cc}
	-\lambda - \mu & 0\\
	0 &  \lambda - \mu
\end{array}\right)
\cdot \textbf{W}_{d}
\end{equation}
where $\lambda$ are the eigenvalues of the Jacobian of the fluid velocity field. If $\mu = \frac{2}{3}St^{-1}$, is very large, then both the components of $A_d$ would decay. For low values of $\mu$, one component of $A_d$ would grow. Therefore the dynamics of an inertial particles depend on the value of $\mu$, that is on the Stokes number. It is reasonable to expect that the computations of the sensitivity of the particles location to the initial relative velocity also would depend on the Stokes number. That this is indeed the case is shown by the computations of the sensitivity field for a particle with Stokes number $0.1$ for the time independent flow, as shown in Figure \ref{particle_separation}(a). The thick lines are the ridges in sensitivity field for particles with $St = 0.1$ and the hatched lines are those of $St = 0.2$. It can be seen that though the structure of the sensitivity field field is similar, the ridges are present at different locations in the relative velocity subspace. This fact can be exploited to design a process to separate particles by their Stokes number. In this section we illustrate a simple procedure for doing this.

The ridges of the sensitivity fields computed for the two different particles of Stokes number 0.1 and 0.2 respectively are overlain in the same plot, as shown in Figure \ref{particle_separation}. 
The subdomain of the relative velocity subspace sandwiched between the ridges of the sensitivity fields of the two types of particles form a zone of segregation. One such sample zone is shown in grey in Figure \ref{particle_separation}. Two particles with $St=0.1$ and $St=0.2$ with initial coordinates $(x,y)= (\frac{3\pi}{8},\frac{3\pi}{8})$ and the same initial relative velocity, belonging to this region, have trajectories that separate in the physical space. To illustrate this, the trajectories of five hundred particles of each Stokes number, starting at the same initial physical point $(x,y)= (\frac{3\pi}{8},\frac{3\pi}{8})$ and with initial relative velocities values belonging to the grey region were computed. The position of these particles is plotted as a function of time to show that the particles are completely segregated into two different cells. 

The above procedure can be applied to other regions sandwiched between the ridges of the partial FTLE of the two different types of particles.

\section{Robustness of the sensitivity field to perturbations in the stream function }

The time-independent flow given the stream function in eq.~\eqref{ti_flow} is perturbed by making it weakly time dependent. The modified fluid flow is given by the stream function
 
\begin{equation}
\psi (x,y,t) = a\cos{(x + b\sin\omega t)} \cos{y}
\end{equation}
The velocity field is given by,

\begin{equation}
u = -a \cos{(x + b\sin\omega t)} \sin{y}
\end{equation}

\begin{equation}
v = a \sin{(x + b\sin\omega t)} \cos{y}
\end{equation}

\begin{figure}[h,t]
\begin{center}
\includegraphics[height=0.34\textwidth]{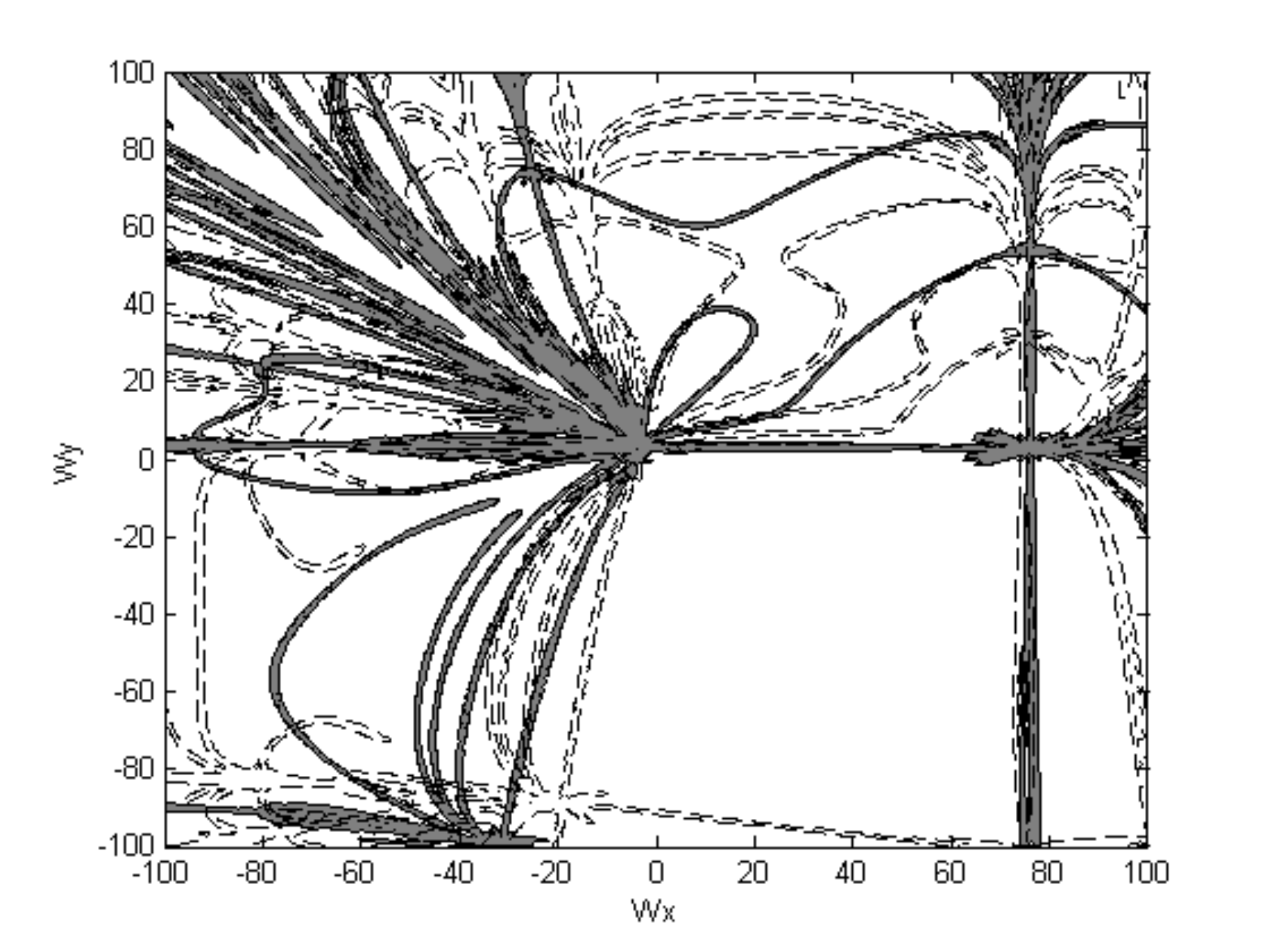} \\{\footnotesize (a)$T_0 = 0.00$}\\
\includegraphics[height =0.34\textwidth]{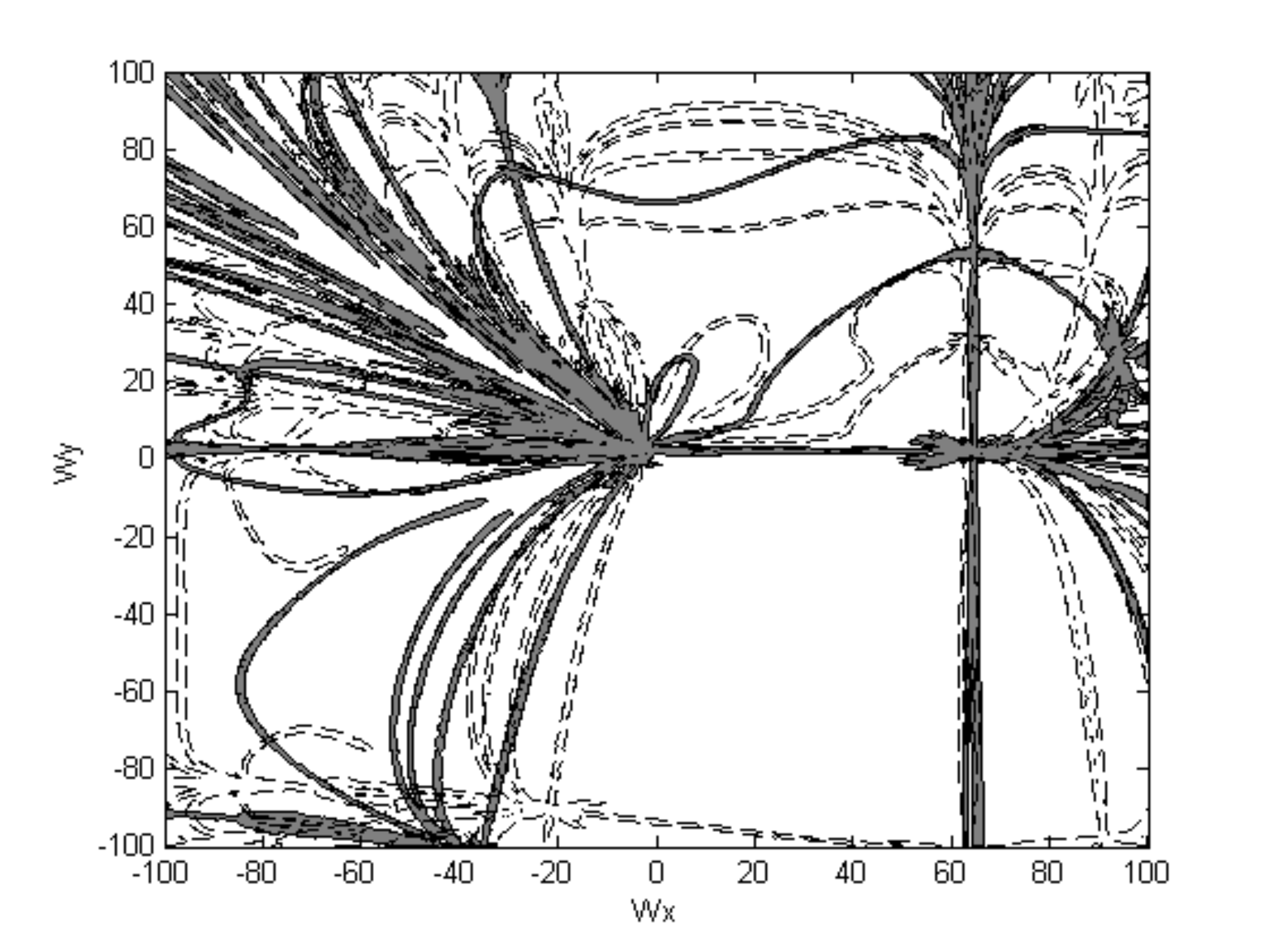} \\{\footnotesize (b)$T_0 = 0.25$}\\
\includegraphics[height =0.34\textwidth]{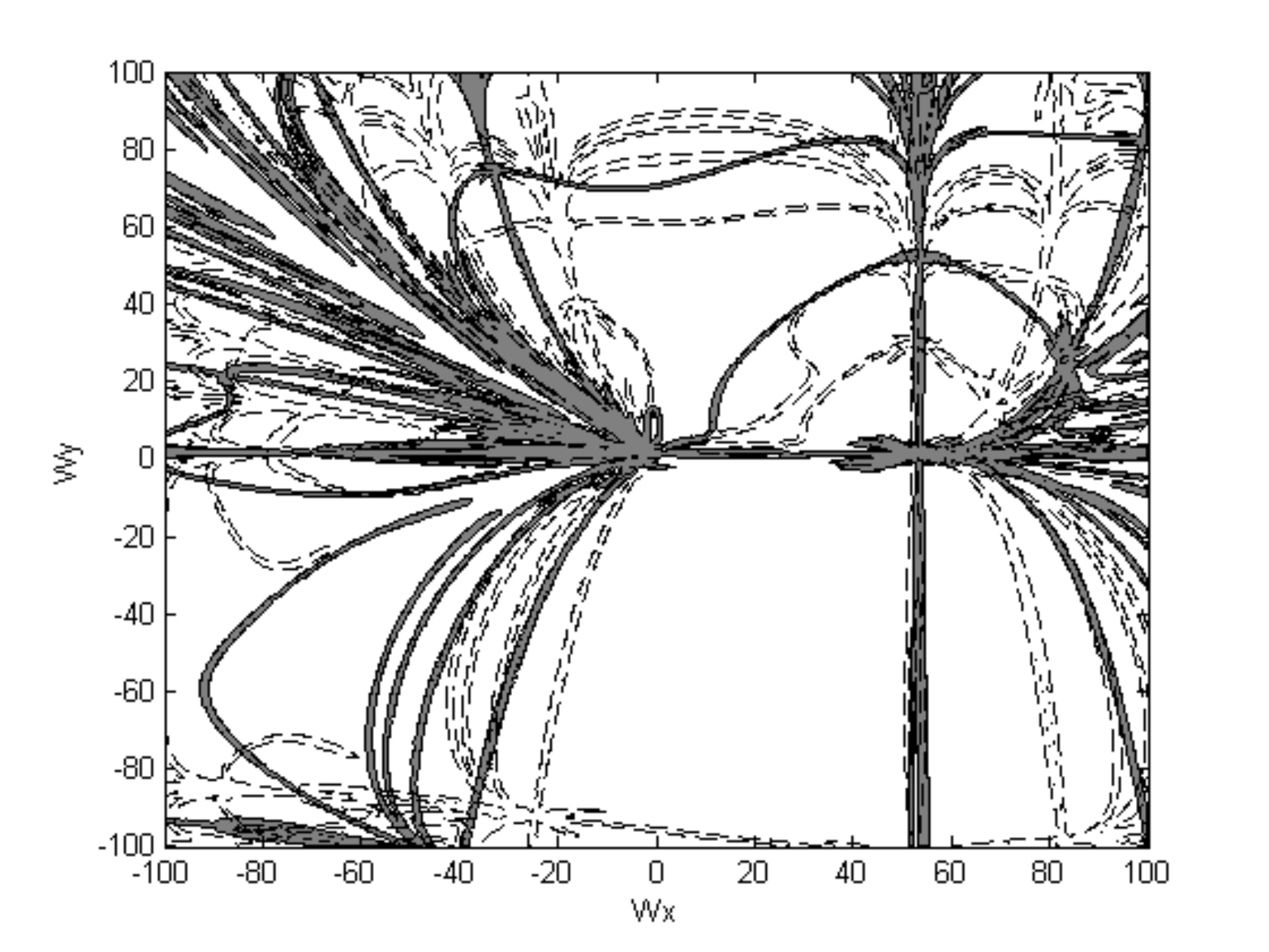} \\{\footnotesize (c)$T_0 = 0.50$}\\
\end{center}\caption{\label{ridge_robust}{\footnotesize
(a) Ridges in the sensitivity field for the stream function $\psi (x,y,t) = a\cos{(x + b\sin\omega t)} \cos{y}$ for $(x_0,y_0) = (3 \pi /8, 3 \pi /8)$. The hatched lines and the thick lines are the ridges corresponding to St = 0.2 and St = 0.1 respectively. Parameters, a = 100, b = 0.25, $ \omega = 1$,Integration time = 0.24. Initial time $T_0 = 0$. 
(b) $T_0 = 0.25$
(c) $T_0 = 0.5$
}}
\end{figure}

For time dependent systems the location of the LCS depends on the choice of initial time. For the computation of the sensitivity field, the location of ridges in the relative velocity subspace depend on the initial spatial coordinates of the particle as well as the initial time. However our computations show that the dependence of the ridge structure on the initial time is weak. Figure \ref{ridge_robust} shows the ridges in the sensitivity field. As the initial time is increased, it is seen that there is a `squeezing' of the sensitivity field in some regions of the relative velocity subspace. A comparison with Figure \ref{particle_separation} and Figure \ref{velocity_partition} shows that the ridge locations in the sensitivity field remain qualitatively the same, for the three cases Figure \ref{ridge_robust} where the initial time is small. This offers a numerical evidence that the sensitivity field is robust to small perturbations in the fluid velocity.

\section{Conclusion }

The dynamics of inertial particles in a fluid flow can exhibit sensitivity to initial conditions. The finite time Lyapunov exponent can be used to characterize this sensitivity. The LCS obtained from the ridges of the FTLE field offers a systematic method to identify qualitatively different regions of the phase space. We demonstrated that even a reduced one dimensional ridge in the sensitivity field contains important information about the sensitivity of the spatial location of particles to initial relative velocity. The Stokes number, and by implication the size of the particle, is an important parameter that governs the clustering behavior of particles for a given flow. This property can be exploited to make particles of different sizes cluster in different regions of the fluid and thus separate them. For the more general case of non neutrally buoyant particles, the density of the particles could play a similar governing role as the Stokes number. One could therefore design flows that can fractionally separate particles for a range of inertial parameters.

\newpage

\newpage

\newpage
\newpage

\newpage

\newpage
\newpage

\end{document}